\newcommand*{\ATLASLATEXPATH}{}
\documentclass[UKenglish,texlive=2011,cernpreprint,txfonts]{atlasdoc}
 \pdfoutput=1
\usepackage[subfigure]{atlaspackage}
\usepackage{multirow}
\usepackage{atlasbiblatex}
\usepackage[process,texmf]{atlasphysics}
\usepackage{SubstructurePerformanceEPJC-defs}

\AtlasTitle{Reconstruction of hadronic decay products of tau leptons with the ATLAS experiment}
\author{The ATLAS Collaboration}

\AtlasRefCode{PERF-2014-06}
\PreprintIdNumber{CERN-PH-EP-2015-294}
\AtlasDate{\today}
\arXivId{1512.05955}

\AtlasJournal{EPJC}

\AtlasAbstract{%
This paper presents a new method of reconstructing the individual charged and
neutral hadrons in tau decays with the ATLAS detector.  The reconstructed
hadrons are used to classify the decay mode and to calculate the visible
four-momentum of reconstructed tau candidates, significantly improving the
resolution with respect to the calibration in the existing tau reconstruction.
The performance of the reconstruction algorithm is optimised and evaluated
using simulation and validated using samples of $Z\to\tau\tau$ and
$Z(\to\mu\mu)$+jets events selected from proton--proton collisions 
at a centre-of-mass energy $\sqrt{s}=8\TeV$, corresponding to an integrated
luminosity of 5~\ifb.  
}

\hypersetup{pdftitle={ATLAS draft},pdfauthor={The ATLAS Collaboration}}

\begin{document}
\maketitle

\section{Introduction}\label{sec:introduction}

\begin{sloppypar}
Final states with hadronically decaying tau leptons play an important part in
the physics programme of the ATLAS experiment~\cite{PERF-2007-01}. Examples
from Run 1 (2009--2013) of the Large Hadron Collider (LHC)~\cite{LHC} are
measurements of Standard Model
processes~\cite{TOPQ-2012-04,TOPQ-2011-05,STDM-2011-23,STDM-2011-46,STDM-2011-18},
Higgs boson searches~\cite{HIGG-2013-32}, including models with extended Higgs
sectors~\cite{HIGG-2012-22,HIGG-2012-09,HIGG-2012-11}, and searches for new
physics phenomena, such as
supersymmetry~\cite{SUSY-2013-10,SUSY-2013-14,SUSY-2012-20}, new heavy gauge
bosons~\cite{EXOT-2012-03} and leptoquarks~\cite{EXOT-2012-14}.  These analyses
depended on robust tau reconstruction and excellent particle identification
algorithms that provided suppression of backgrounds from jets, electrons and
muons~\cite{Run1TauPaper}. 

With the discovery of a Higgs boson~\cite{HIGG-2012-27,CMS-HIG-12-028} and
evidence for the Higgs-boson Yukawa coupling to tau
leptons~\cite{HIGG-2013-32,CMS-HIG-13-004}, a key future measurement will be
that of the $CP$ mixture of the Higgs boson via spin effects in $H\to\tau\tau$
decays~\cite{CPDesch,CPHarnik,CPBerge}. This measurement relies on high-purity
selection of the $\Ptauon\to\piminus\nu$, $\Ptauon\to\piminus\pizero\nu$ and
$\Ptauon\to\piminus\piplus\piminus\nu$ decays, as well as the reconstruction of
the individual charged and neutral pion four-momenta. The tau reconstruction
used in ATLAS throughout Run 1 (here denoted as ``Baseline''), however, only
differentiates tau decay modes by the number of charged hadrons and does not
provide access to reconstructed neutral pions. 

This paper presents a new method (called ``\newmethod'') of reconstructing the
individual charged and neutral hadrons in tau decays with the ATLAS detector.
Charged hadrons are reconstructed from their tracks in the tracking system.
Neutral pions are reconstructed from their energy deposits in the calorimeter.
The reconstructed hadrons, which make up the visible part of the tau decay
(\tauvis), are used to classify the decay mode and to calculate the
four-momentum of reconstructed \tauvis candidates. The superior four-momentum
resolution from the tracking system compared to the calorimeter, for charged
hadrons with transverse momentum (\pt) less than $\sim$$100\GeV$, leads to a
significant improvement in the tau energy and directional resolution. This
improvement, coupled with the ability to better identify the hadronic tau decay
modes, could lead to better resolution of the ditau mass
reconstruction~\cite{MMC}. The performance of the \newmethod is validated using
samples of real hadronic tau decays and jets in $Z$+jets events selected from
data. The samples correspond to $5\,\ifb$ of data collected during
proton--proton collisions at a centre-of-mass energy of $\sqrt{s} = 8\TeV$,
which was the amount of data reprocessed using \newmethod.  While similar
concepts for the reconstruction of hadronic tau decays have been employed at
other
experiments~\cite{Schael:2005am,Akers:1994ex,Abdallah:2003cw,Acciarri:1998vg,CDFPFlow,D0Ztautau,Run1TauCMS},
the \newmethod is specifically designed to exploit the features of the ATLAS
detector and to perform well in the environment of the LHC.

The paper is structured as follows.  The ATLAS detector, event samples, and
the reconstruction of physics objects used to select \tauhadvis candidates from
the $8\TeV$ data are described in Section~\ref{sec:detector}.
The properties of \tauhadvis decays and the \newmethod method are described
in Section~\ref{sec:reco}, including its concepts (Section~\ref{sec:concepts}),
neutral pion reconstruction (Section~\ref{sec:reconeutrals}), reconstruction of
individual photon energy deposits (Section~\ref{sec:recoshots}), decay mode
classification (Section~\ref{sec:recodecaymode}) and \tauhadvis four-momentum
reconstruction (Section~\ref{sec:fourmomentum}).  Conclusions are presented 
in Section~\ref{sec:conclusions}.
\end{sloppypar}

\section{ATLAS detector and event samples}\label{sec:detector}

\subsection{The ATLAS detector}
\label{sec:atlas}

The ATLAS detector~\cite{PERF-2007-01} consists of an inner tracking
system surrounded by a superconducting solenoid, electromagnetic (EM)
and hadronic (HAD) calorimeters, and a muon spectrometer. The
inner detector is immersed in a 2 T axial magnetic field, and
consists of pixel and silicon microstrip detectors inside a
transition radiation tracker, which together provide charged-particle
tracking in the region $|\eta|<2.5$.\footnote{ATLAS uses a
  right-handed coordinate system with its origin at the nominal
  interaction point (IP) in the centre of the detector and the
  $z$-axis along the beam direction. The $x$-axis points from the IP
  to the centre of the LHC ring, and the $y$-axis points
  upward. Cylindrical coordinates $(r,\phi)$ are used in the
  transverse $(x,y)$ plane, $\phi$ being the azimuthal angle around
  the beam direction. The pseudorapidity is defined in terms of the
  polar angle $\theta$ as $\eta=-\ln\tan(\theta/2)$.  The distance
  $\Delta{}R$ in the $\eta$--$\phi$ space is defined as
  $\Delta{}R=\sqrt{({\Delta\eta})^2 + ({\Delta\phi})^2}$.}  
The EM calorimeter is based on lead and liquid argon as absorber and active
material, respectively. In the central rapidity region, the EM calorimeter is
divided radially into three layers: the innermost layer (EM1) is finely
segmented in $\eta$ for optimal $\gamma / \pi^0$ separation, the layer next in
radius (EM2) collects most of the energy deposited by electron and photon
showers, and the third layer (EM3) is used to correct leakage beyond the EM
calorimeter for high-energy showers.  A thin presampler layer (PS) in front of
EM1 and in the range $|\eta| < 1.8$ is used to correct showers for upstream
energy loss.  Hadron calorimetry is based on different detector technologies,
with scintillator tiles ($|\eta| < 1.7$) or liquid argon ($1.5 < |\eta| < 4.9$)
as active media, and with steel, copper, or tungsten as absorber material. The
calorimeters provide coverage within $|\eta|<4.9$.  The muon spectrometer
consists of superconducting air-core toroids, a system of trigger chambers
covering the range $|\eta|<2.4$, and high-precision tracking chambers allowing
muon momentum measurements within $|\eta|<2.7$.  A three-level trigger system
is used to select interesting events~\cite{PERF-2011-02}. The first-level
trigger is implemented in hardware and uses a subset of detector information to
reduce the event rate to a design value of at most 75\,kHz. This is followed by
two software-based trigger levels which together reduce the average event rate
to 400\,Hz.

\subsection{Physics objects}
\label{sec:physicsobjects}

This section describes the \baseline \tauhadvis reconstruction and also the
reconstruction of muons and the missing transverse momentum, which are required
for the selection of samples from data. \newmethod operates on each
reconstructed \baseline tau candidate to reconstruct the charged and neutral
hadrons, classify the decay mode and to provide an alternative \tauhadvis
four-momentum. Suppression of backgrounds from other particles misidentified as
\tauhadvis is achieved independently of the \newmethod. 

The \baseline \tauhadvis reconstruction and energy calibration, and the
algorithms used to suppress backgrounds from jets, electrons and muons are
described in detail in Ref.~\cite{Run1TauPaper}.  Candidates for hadronic tau
decays are built from jets reconstructed using the \mbox{anti-$k_{t}$}
algorithm~\cite{antikt,antikt-fast} with a radius parameter value of 0.4.
Three-dimens\-ional clusters of calorimeter cells calibrated using a local
hadronic calibration~\cite{ATL-LARG-PUB-2008-002,LCCalibration} serve as inputs
to the jet algorithm. The calculation of the \tauhadvis{} four-momentum uses
clusters within the {\em core region} ($\Delta R<0.2$ from the initial
jet-axis). It includes a final tau-specific calibration derived from simulated
samples, which accounts for out-of-cone energy, underlying event, the typical
composition of hadrons in hadronic tau decays and contributions from multiple
interactions occurring in the same and neighbouring bunch crossings (called
\pileup). Tracks reconstructed in the inner detector are matched to the
\tauhadvis{} candidate if they are in the core region and satisfy the following
criteria: $\pt>1\GeV$, at least two associated hits in the pixel layers of the
inner detector, and at least seven hits in total in the pixel and silicon
microstrip layers. Furthermore, requirements are imposed on the distance of
closest approach of the tracks to the tau primary vertex in the transverse
plane, $|d_0|<1.0$\,mm, and longitudinally, $|z_0\sin\theta|<1.5$\,mm.  The
\tauhadvis{} charge is reconstructed from the sum of the charges of the
associated tracks. 

Backgrounds for \tauhadvis candidates originating from quark- and
gluon-initiated jets are discriminated against by combining shower shape and
tracking information in a multivariate algorithm that employs boosted decision
trees (BDTs)~\cite{TMVA}.  The efficiency of the jet discrimination algorithm
has little dependence on the \pt{} of the \tauhadvis candidates (evaluated
using candidates with $\pt>15\GeV$) or on the number of reconstructed primary
vertices, which is correlated to the amount of \pileup, and has been evaluated
up to a maximum of 25 primary vertices per event.  All \tauhadvis candidates
are required to have \mbox{$\pt{} > 15\GeV$}, to be in the fiducial volume of
the inner detector, \mbox{$|\eta| < 2.5$}, and to have one or three associated
tracks.  They must also meet jet discrimination criteria, corresponding to an
efficiency of about 55\% (40\%) for simulated \tauhadvis with one (three)
charged decay products~\cite{Run1TauPaper}, leading to a rate of false
identification for quark- and gluon-initiated jets of below a percent. 
A discriminant designed to suppress candidates arising from the
misidentification of electrons~\cite{Run1TauPaper} is also applied.

Muons are reconstructed using tracks in the muon spectrometer and inner
detector~\cite{PERF-2014-05}.  The missing transverse momentum is computed from
the combination of all reconstructed and fully calibrated physics objects and
the remaining clustered energy deposits in the calorimeter not associated with
those objects~\cite{PERF-2011-07}.

\subsection{Event samples and selection}\label{sec:samples}

The optimisation and measurement of the \tauvis reconstruction performance
requires Monte Carlo simulated events. Samples of simulated $pp$ collision
events at $\sqrt{s}=8\TeV$ are summarised in Table~\ref{tab:samples}.  Tau
decays are provided by \Ztautau events.  The {\em sophisticated tau decay}
option of \pythiaeight{} is used, which provides fully modelled hadronic decays
with spin correlations~\cite{Pythia8Taus}. Tau decays in the \ttbar sample are
generated by \tauola{}~\cite{TAUOLA}. Photon radiation is performed by
\photos{}~\cite{PHOTOS}.  Single-pion samples are also used, in which the pions
originate from the centre of the ATLAS detector and are generated to have a
uniform distribution in $\phi$ and $\eta$ ($|\eta|<5.5$) and also in $\log(E)$
($200\MeV < E < 2\TeV$). 

\begin{table}[tbp]
\centering
\begin{tabular}{lllll}
\toprule
Process                   & Generator                                  & PDFs                   & UE tune                               \\
\midrule
\Ztautau                  & \pythiaeight~\cite{Pythia8}                & \cteq~\cite{CTEQ6L1}   & \autwo{}~\cite{ATL-PHYS-PUB-2012-003} \\ 
\Wmunu                    & \alpgen~\cite{Alpgen}+\pythiaeight         & \cteq                  & \perugia{}~\cite{Perugia}             \\
\Wtaunu                   & \alpgen+\pythiaeight                       & \cteq                  & \perugia                              \\
\Zmm                      & \alpgen+\pythiaeight                       & \cteq                  & \perugia                              \\
\ttbar                    & \mcnlo~\cite{MCNLO,MCNLO1,MCNLO2}+\herwig{}~\cite{Herwig6,Jimmy}          & \ctten~\cite{CT10}     & \auettwo{}~\cite{ATL-PHYS-PUB-2012-003} \\ 
\bottomrule
\end{tabular}
\caption{
    Details regarding the simulated samples of $pp$ collision events. The following
    information is provided for each sample: the generator of the hard interaction,
    parton shower, hadronisation and multiple parton interactions; the set of
    parton distribution functions (PDFs) and the underlying event (UE) tune of the Monte Carlo. 
}
\label{tab:samples}
\end{table}

The response of the ATLAS detector is simulated using
\geant~\cite{SOFT-2010-01,geant4} with the hadronic-shower model
\qgspbert~\cite{QGS,Bertini}. The parameters of the underlying event (UE)
simulation were tuned using collision data. Simulated $pp$ collision events are
overlaid with additional minimum-bias events generated with \pythiaeight{} to
account for the effect of \pileup.  When comparing to the data, the simulated
events are reweighted so that the distribution of the number of \pileup
interactions matches that in the data.  The simulated events are reconstructed
with the same algorithm chain as used for the collision data. 

Samples of \tauhadvis{} candidates are selected from the data using a {\it
tag-and-probe} approach.  Candidates originating from hadronic tau decays and
jets are obtained by selecting \Ztautau and \Zmumujets  events, respectively.
The data were collected by the ATLAS detector during $pp$ collisions at
\mbox{$\sqrt{s}=8\TeV$}. The sample corresponds to an integrated luminosity of
\lumi after making suitable data quality requirements for the operation of the
tracking, calorimeter, and muon spectrometer subsystems. The data have a
maximum instantaneous luminosity of $7\cdot10^{33}$\,cm$^{-2}$\,s$^{-1}$ and an
average number of 19 $pp$ interactions in the same bunch crossing.  

The \Ztautau tag-and-probe approach follows Ref.~\cite{Run1TauPaper}; events
are triggered by the presence of a muon from a leptonic tau decay ({\it tag})
and must contain a \tauhadvis candidate ({\it probe}) with \mbox{$\pt{} > 20\GeV$}, which is used to
evaluate the tau reconstruction performance. The \tauhadvis selection criteria
described in Section~\ref{sec:physicsobjects} are used. In addition the
\tauhadvis must have unit charge which is opposite to that of the muon.
A discriminant designed to suppress candidates arising from the
misidentification of muons~\cite{Run1TauPaper} is also applied
to increase signal purity. The invariant mass of the muon and \tauhadvis,
$m(\mu,\tauhadvis)$, is required to be in the range $50\GeV <
m(\mu,\tauhadvis) < 85\GeV$, as expected for \Ztautau decays.  The background
is dominated by multijet and \Wmunujets{} production and is estimated using the
techniques from Ref.~\cite{STDM-2011-18}.

The \Zmumujets tag-and-probe approach follows Ref.~\cite{STDM-2012-04}, with
the following differences: both muons are required to have $\pt>26$~\GeV, the dimuon
invariant mass must be between 81 and 101~\GeV, and the highest-\pt jet is
selected as a probe \tauhadvis candidate if it satisfies the \tauhadvis selection
criteria described in Section~\ref{sec:physicsobjects} but with \mbox{$\pt{} > 20\GeV$} and without the electron discriminant. In this approach, two more 
steps are made when comparing simulated events to the data. Before the \tauhadvis
selection, the simulated events are reweighted so that the \pt distribution of
the $Z$ boson matches that in data. After the full event selection, the overall
normalisation of the simulation is scaled to that in the data.

\section{Reconstruction of the \tauhadvis}
\label{sec:reco}

Over 90\% of hadronic tau decays occur through just five dominant decay modes,
which yield one or three charged hadrons (\charged), up to two neutral pions
(\pizero) and a tau neutrino. The neutrino goes undetected and is omitted in
further discussion of the decay modes. Table~\ref{tab:decaymodes} gives the
following details for each of the five decay modes: the branching fraction,
$\mathcal{B}$; the fraction of simulated \tauhadvis candidates that pass the
\tauhadvis selection described in Section~\ref{sec:physicsobjects}  without the
jet and electron discrimination, $\mathcal{A}\cdot\varepsilon_{\rm reco}$; and the fraction
of those that also pass the jet and electron discrimination, $\varepsilon_{\rm ID}$. The
\chargeds are predominantly $\pipm$'s with a minor contribution from
$K^{\pm}$'s. The modes with two or three pions proceed mainly through the
intermediate $\rho$ or $a_1$ resonances, respectively. The \chargeds are
sufficiently long-lived that they typically interact with the detector before
decaying and are therefore considered stable in the \newmethod.  The \pizeros
decay almost exclusively to a pair of photons.  Approximately half of the
photons convert into an $e^+e^-$ pair because of interactions with the beampipe
or inner-detector material. 
Modes with more \pizeros tend to have lower $\varepsilon_{\rm ID}$ as they have 
wider showers that are more similar to those produced by quark- and gluon-initiated
jets. The mode dependence of $\mathcal{A}\cdot\varepsilon_{\rm reco}$ is due to 
a mixture of effects. The fraction of energy carried by visible decay products 
is mode dependent and the response of the calorimeter to \chargeds and \pizeros is different, 
both of which impact the efficiency of the \tauhadvis \pt requirement. 
The efficiency of the track association is also dependent on the number of \chargeds
and to a lesser extent the number of \pizeros, which can contribute tracks from 
conversion electrons.  

The goal of the \newmethod is to classify the five decay modes and to
reconstruct the individual \chargeds and \pizeros. The performance is evaluated
using  the energy and directional residuals of \pizero and \tauhadvis and the
efficiency of the \tauhadvis decay mode classification. The $\eta$ and $\phi$
residuals are defined with respect to the generated values: $\eta - \eta^{\rm
gen}$ and $\phi - \phi^{\rm gen}$, respectively. For $\et$, the {\em relative}
residual is defined with respect to the generated value $\et/\et^{\rm gen}$.  
The {\em core} and {\em tail} resolutions for $\eta$, $\phi$ and $\et$
are defined as half of the 68\% and 95\% central intervals of their 
residuals, respectively.
Decays into higher-multiplicity states are accommodated by including modes with
more than two \pizeros in the \onepxn category and more than one \pizero in the
\threepxn category.  Decays with more than three charged hadrons are not
considered. No attempt is made to reconstruct neutral kaons or to separate
charged kaons from charged pions.

\begin{table}[tbp]
\centering
\begin{tabular}{lccc}
\toprule
Decay mode & $\mathcal{B}$ [\%] & $\mathcal{A}\cdot\varepsilon_{\rm reco}$ [\%] & $\varepsilon_{\rm ID}$ [\%] \\
\midrule
$\onepzeron$   &  11.5  & 32 & 75 \\
$\oneponen$    &  30.0  & 33 & 55 \\
$\onepxn$      &  10.6  & 43 & 40 \\
$\threepzeron$ &   9.5  & 38 & 70 \\
$\threepxn$    &   5.1  & 38 & 46 \\
\bottomrule
\end{tabular}
\caption{
Five dominant \tauhadvis decay modes~\cite{PDG2014}. Tau neutrinos are omitted from the table. The symbol $h^\pm$ stands
for $\pipm$ or $K^\pm$. Decays involving $K^\pm$ contribute $\sim$3\% to the
total hadronic branching fraction. Decays involving neutral kaons are excluded.
The branching fraction ($\mathcal{B}$), the fraction of generated \tauhadviss in simulated \Ztautau events that are
reconstructed and pass the \tauhadvis selection described in
Section~\ref{sec:physicsobjects} without the jet and electron discrimination
($\mathcal{A}\cdot\varepsilon_{\rm reco}$) and the fraction of those \tauhadvis candidates
that also pass the jet and electron discrimination ($\varepsilon_{\rm ID}$) for each decay
mode are given.
}
\label{tab:decaymodes}
\end{table}

\subsection{Concepts of the \newmethod method}
\label{sec:concepts}

The main focus of the \newmethod method is to reconstruct \tauhadviss with
\pt{} values between 15~and $100\GeV$, which is the relevant range for tau leptons
produced in decays of electroweak and SM Higgs bosons. In this case the hadrons
typically have \pt{} lower than $20\GeV$ (peaked at $\sim$$4\GeV$) and have an
average separation of $\Delta R\approx0.07$. The \chargeds are
reconstructed using the tracking system, from which the charge and momentum
are determined. Each track associated with the \tauhadvis candidate in
the core region is considered to be a \charged and the \pipm mass hypothesis is
applied. 
Approximately 2\% of the selected \tauhadviss have a misclassified number of \chargeds.
Overestimation of the number of \chargeds is primarily due to
additional tracks from conversion electrons, which are highly suppressed by the
strict track selection criteria described in Section~\ref{sec:physicsobjects}.
Underestimation of the number of \chargeds is primarily caused by tracking
inefficiencies ($\sim$10\% for charged pions with
\mbox{$\pt>1\GeV$}~\cite{PERF-2007-01}), which arise from interactions of the
\chargeds with the beampipe or detector material.
The \chargeds also produce a shower in the calorimeter from which their energy and
direction can be determined, but the tracker has a better performance in
the relevant momentum range. The shower shapes of \chargeds are also highly
irregular, with a typical width of \mbox{$0.02<\Delta R<0.07$} in the EM calorimeter,
combined with large fluctuations in the fractional energy depositions in the
layers of the calorimeter. The \pizeros are reconstructed from their energy
deposits in the EM calorimeter. The main challenge is to disentangle their
energy deposits from \charged showers, which have a width similar to the
average separation between hadrons. The photons from \pizero decays are highly collimated, with a
typical separation of \mbox{$0.01<\Delta R < 0.03$}. 
The majority of the \pizero energy is reconstructed in a single cluster in the
EM calorimeter. Compared to \chargeds, \pizero showers are smaller and more
regular, leaving on average 10\%, 30\% and 60\% of their energy in PS, EM1 and EM2,
respectively. Almost no \pizero energy is deposited beyond EM2, so EM3 is
considered part of the HAD calorimeter in \newmethod.  The characteristic
shower shapes and the kinematics of \chargeds and \pizeros  are used to identify
\pizeros and to classify the tau decay mode. 

\begin{sloppypar} 
In the following sections, the individual steps of the \newmethod method for
\tauhadvis reconstruction are described. The first step is the reconstruction
and identification of neutral pions. Next, energy deposits from individual
photons in the finely segmented EM1 layer are reconstructed to identify cases
where two \pizeros are contained within a single cluster. The decay mode is
then classified by exploiting the available information from the reconstructed
\chargeds and \pizeros and the photons reconstructed in EM1. Following the
decay mode classification, the \tauhadvis four-momentum is reconstructed from
the individual hadrons and then combined with the \baseline energy calibration
to reduce tails in the \et residual distribution.  The performance of the
\newmethod is evaluated using \tauhadvis candidates from simulated \Ztautau
events.
\end{sloppypar}

\subsection{Reconstruction and identification of neutral pions}
\label{sec:reconeutrals}

The reconstruction of neutral pion candidates (\neutral) within hadronic tau
decays using the \newmethod proceeds as follows. First, \neutrals are created
by clustering cells in the EM calorimeter in the core region of the \tauhadvis.
In the next step, the \neutral energy is corrected for contamination from
\chargeds. To do this, the energy that each \charged deposits in the EM
calorimeter (\EpionEM) is estimated as the difference between the energy of the
\charged from the tracking system (\ppiontrk{}) and the energy deposited in the
HAD calorimeter which is associated with the \charged (\EpionHAD{}): $\EpionEM
= \ppiontrk - \EpionHAD$.  To calculate \EpionHAD{}, all clustered energy
deposits in the HAD calorimeter in the core region are assigned to the closest
\charged, determined using the track position extrapolated to the calorimeter
layer that contains most of the cluster energy. The \EpionEM{} of each \charged
is then subtracted from the energy of the closest \neutral if it is within
\mbox{$\dR=0.04$} of the \charged.

\begin{figure}[htbp]
\begin{center}
\subfigure[\label{subfig:pionIDvar}]{\includegraphics[width=0.49\columnwidth]{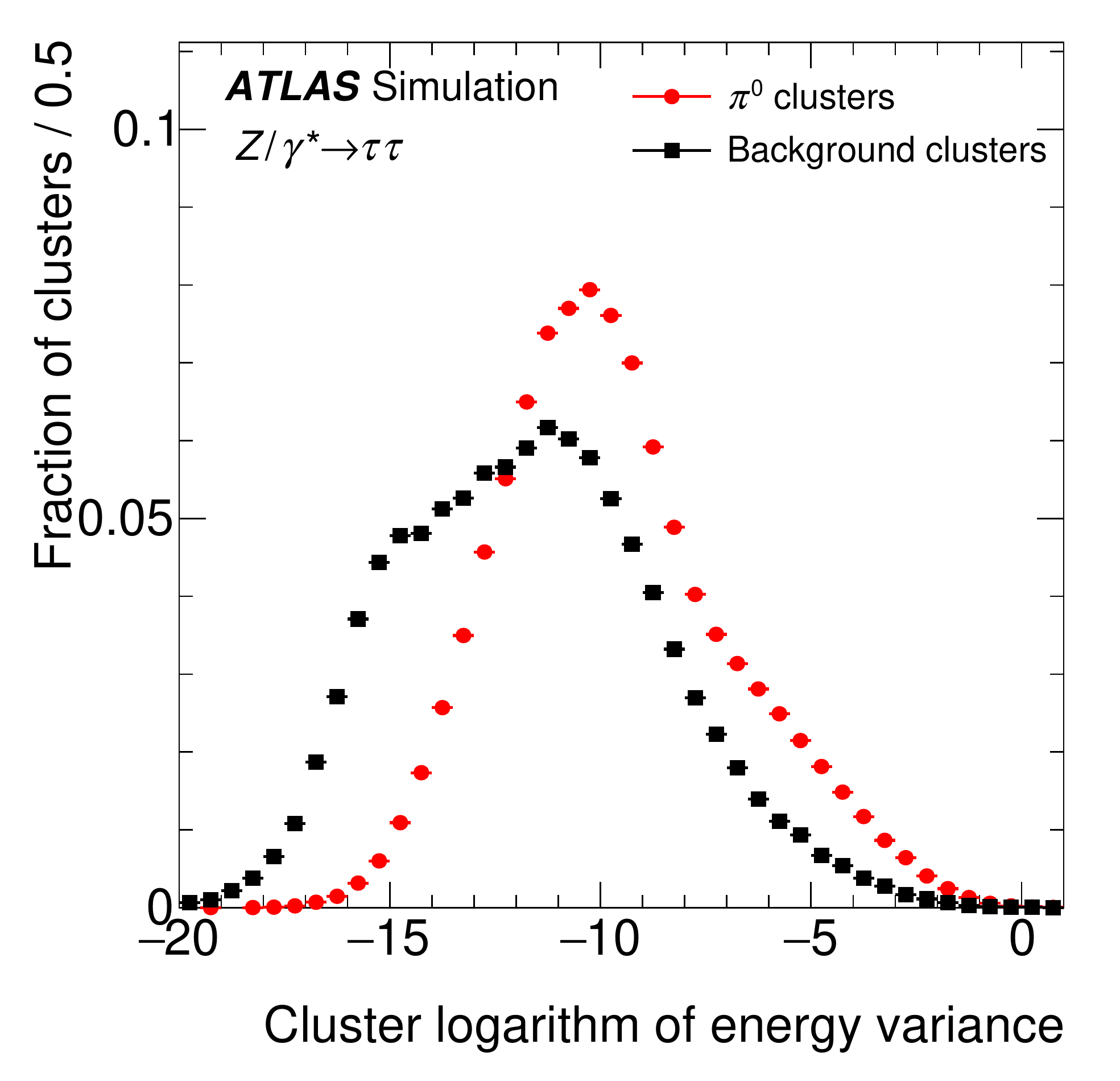}}
\subfigure[\label{subfig:pionROC}]{\includegraphics[width=0.49\columnwidth]{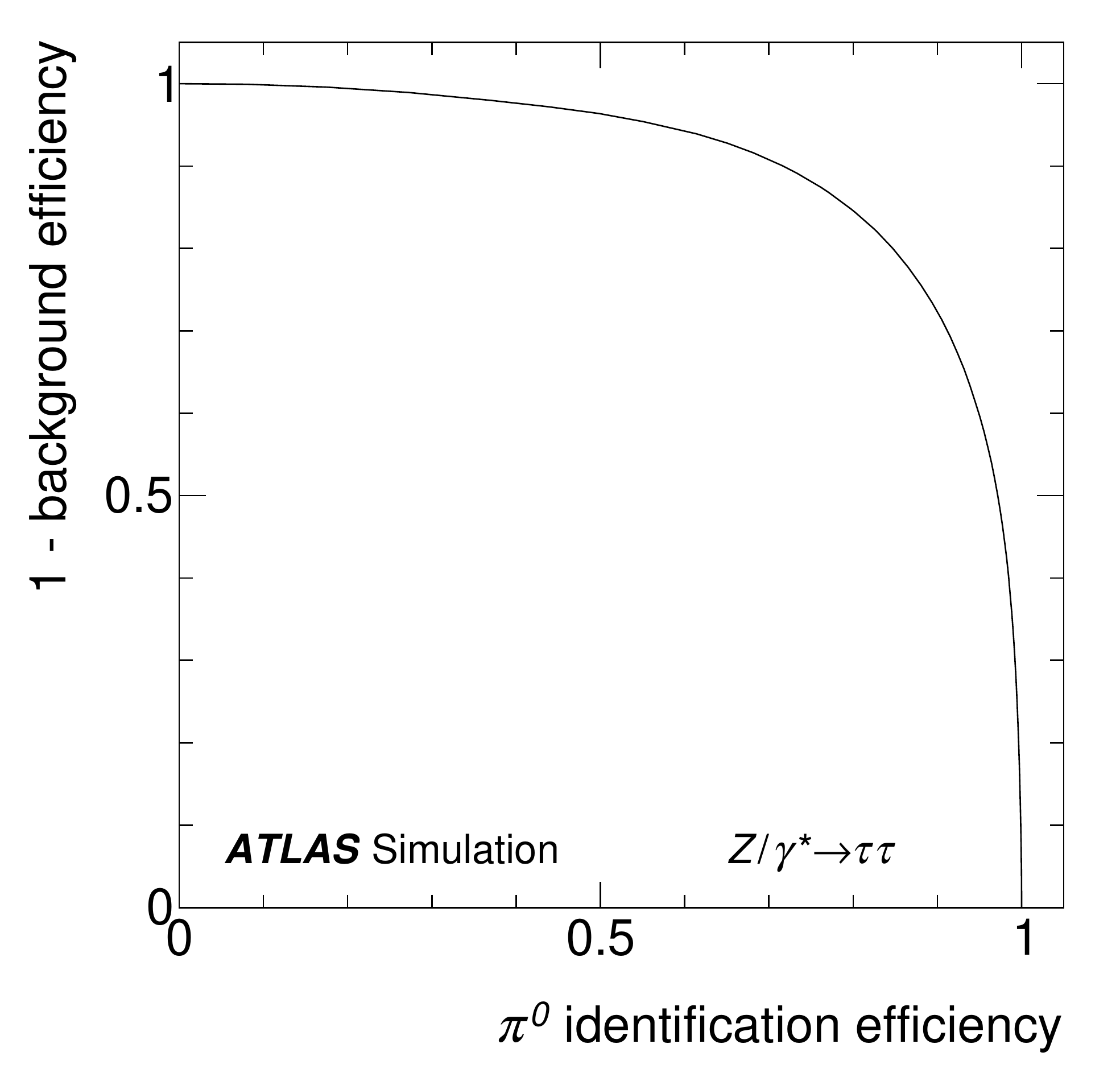}}
\caption{ 
(a) Distribution of the logarithm of the second moment in energy density of \neutral clusters that 
do (signal) or do not (background) originate from \pizeros, as used in the 
\pizero identification.
(b) 1 $-$ efficiency for background \neutrals vs. the efficiency for signal \neutrals 
to pass thresholds on the \pizero identification score. 
The \neutrals in both figures are associated with \tauhadviss selected from simulated \Ztautau events. 
}\label{fig:pionROC}
\end{center}
\end{figure}

At this stage, many of the \neutrals in reconstructed hadronic tau decays do
not actually originate from \pizeros, but rather from \charged remnants,
\pileup or other sources. The purity of \neutrals is improved by applying a
minimum \pt{} requirement and an identification criterion designed to reject
\neutrals not from \pizeros. The \pt{} thresholds are in the range 2.1--$2.7\GeV$.
After the \pt requirement the background is
dominated by \charged remnants.  The \pizero identification uses a BDT and
exploits the properties of the \neutral clusters, such as the energy density
and the width and depth of the shower.  The variables used for \neutral
identification are described in Table~\ref{tab:pi0idvars}.  
The BDT is trained using \tauhadviss that have only one \charged, and which are
produced in simulated \Ztautau{} events.  The \neutrals are assigned to signal
or background based on whether or not they originated from a generated \pizero.
Figure~\ref{subfig:pionIDvar} shows signal and background distributions for the
logarithm of the second moment in energy density, which is one of the more
important identification variables.  The discriminating power of the \pizero
identification is quantified by comparing the efficiency of signal and
background \neutrals to pass thresholds on the identification score, as shown
in Fig.~\ref{subfig:pionROC}.  The \pt{} and identification score thresholds
are optimised in five $|\eta|$ ranges, corresponding to structurally different
regions of the calorimeter, to maximise the number of \tauhadviss with the
correct number of reconstructed \chargeds and identified
\neutrals~(\neutralIDs).

\begin{figure}[htbp]
\begin{center}
\includegraphics[width=0.5\columnwidth]{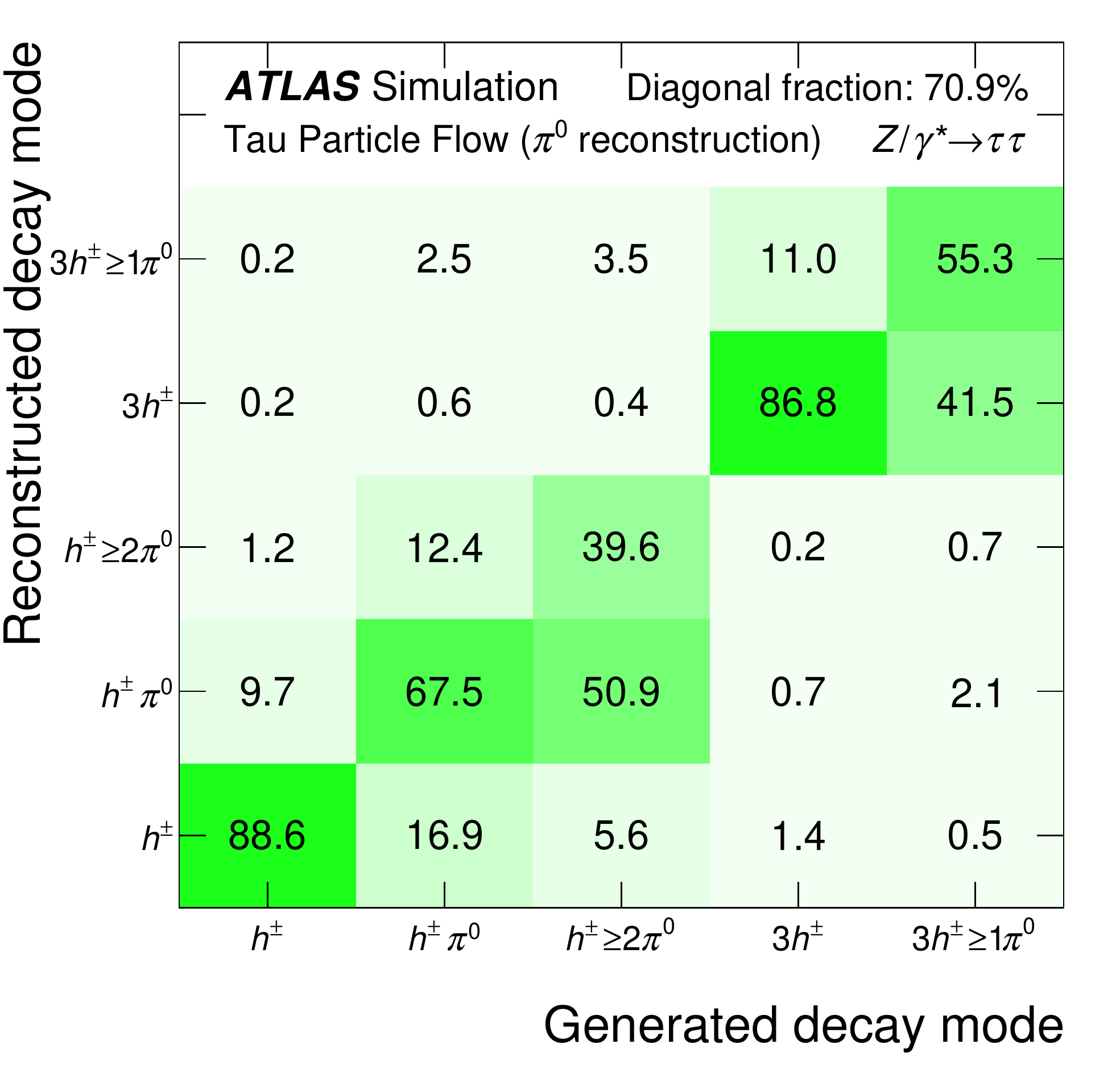}
\caption{
Decay mode classification efficiency matrix showing the probability 
for a given generated mode to be reconstructed as a particular mode
by the \newmethod after \pizero reconstruction in simulated \Ztautau{} 
events. Decays containing neutral kaons are omitted. Only decays from 
\tauhadviss that are reconstructed and pass the selection described in 
Section~\ref{sec:physicsobjects} are considered. The
statistical uncertainty is negligible.  
}
\label{fig:cellbasedMigrationMatrix}
\end{center}
\end{figure}

The \charged and \pizero counting performance is depicted in
Fig.~\ref{fig:cellbasedMigrationMatrix} by a {\em decay mode classification
matrix} which shows the probability for a given generated mode to be
reconstructed as a particular mode. Only \tauhadvis decays that are
reconstructed and pass the selection described in
Section~\ref{sec:physicsobjects} are considered (corresponding efficiencies are
given in Table~\ref{tab:decaymodes}). 
The total fraction of correctly classified tau decays (diagonal fraction) is 70.9\%.
As can be seen, for \tauhadviss with one
\charged, the separation of modes with and without \pizeros is quite good, but
it is difficult to distinguish between \oneponen and \onepxn.  The largest
contributions to the misclassification arise from \onepxn decays where one of
the \pizeros failed selection or where the energy deposits of both \pizeros
merge into a single cluster.  It is also difficult to distinguish between the
\threepzeron and \threepxn modes because the \pizeros are typically soft with
large overlapping \charged deposits.

\begin{table}
\centering
\begin{tabular}{p{0.80\textwidth}}
\toprule 

{\bf Cluster pseudorapidity, \clusterEta} \\ 
Magnitude of the energy-weighted $\eta$ position of the cluster \\[1ex]

{\bf Cluster width,  \clusterWidth} \\
Second moment in distance to the shower axis \\[1ex]

{\bf Cluster $\eta$ width in EM1,  \clusterEtaWidthEMOne} \\
Second moment in $\eta$ in EM1 \\[1ex] 

{\bf Cluster $\eta$ width in EM2,  \clusterEtaWidthEMTwo} \\
Second moment in $\eta$ in EM2 \\[1ex] 

{\bf Cluster depth,  \clusterDepth}\\
Distance of the shower centre from the calorimeter front face measured along the shower axis \\[1ex]

{\bf Cluster PS energy fraction,  \clusterPSFrac}\\
Fraction of energy in the PS \\[1ex]

{\bf Cluster core energy fraction,  \clusterHotCellFrac}\\
Sum of the highest cell energy in PS, EM1 and EM2 divided by the total energy\\[1ex]

{\bf Cluster logarithm of energy variance,  \clusterEnergyVariance}\\
Logarithm of the second moment in energy density \\[1ex]

{\bf Cluster EM1 core energy fraction,  \clusterEMOneCoreFrac}\\
Energy in the three innermost EM1 cells divided by the total energy in EM1\\[1ex]

{\bf Cluster asymmetry with respect to track,  \clusterAsymWRTTrk}\\
Asymmetry in $\eta$--$\phi$ space of the energy distribution in EM1 with respect to the extrapolated track position\\[1ex]

{\bf Cluster EM1 cells,  \clusterNCellsEMOne}\\
Number of cells in EM1 with positive energy\\[1ex] 

{\bf Cluster EM2 cells,  \clusterNCellsEMTwo}\\
Number of cells in EM2 with positive energy\\[1ex]

\bottomrule

\end{tabular}
\caption{
Cluster variables used for \neutral identification.  The variables
\clusterEta{}, \clusterWidth{}, \clusterDepth{}, \clusterHotCellFrac{} and
\clusterEnergyVariance{} are taken directly from the cluster
reconstruction~\cite{LCCalibration}. To avoid confusion with other
variables used in tau reconstruction, the superscript {\em clus} has been added
to each variable. 
}
\label{tab:pi0idvars}
\end{table}

\begin{sloppypar} 
Two alternative methods for \pizero reconstruction were also developed.  In the
first method (\clusterbased) the number of \pizeros in the core region is first
estimated from global tau features measured using calorimetric quantities and
the momenta of the associated \charged tracks. Clusters in the EM calorimeter
are then chosen as \neutrals using a \pizero likeness score based on their
energy deposition in the calorimeter layers and the \tauhadvis track momenta.
The likeness score does not exploit cluster moments to the same extent as the
\pizero identification of the \newmethod and cluster moments are not used at
all to estimate the number of \pizero.  This method was used to calculate
variables for jet discrimination in Run 1~\cite{Run1TauPaper}, but was not
exploited further.  
The other method ({\em shower shape subtraction}, \cellbased) is a modified
version of \newmethod, which attempts to subtract the \charged shower from the
calorimeter at cell level using average shower shapes derived from simulation.
The shower shapes are normalised such that their integral corresponds to
\EpionEM{} and centred on the extrapolated position of the \charged track. They
are then subtracted from the EM calorimeter prior to the clustering, replacing
the cluster-level subtraction of \EpionEM{}. 

\begin{figure}[htbp]
\begin{center}
\subfigure[\label{subfig:neutralpion_resoet}]{\includegraphics[width=0.45\columnwidth]{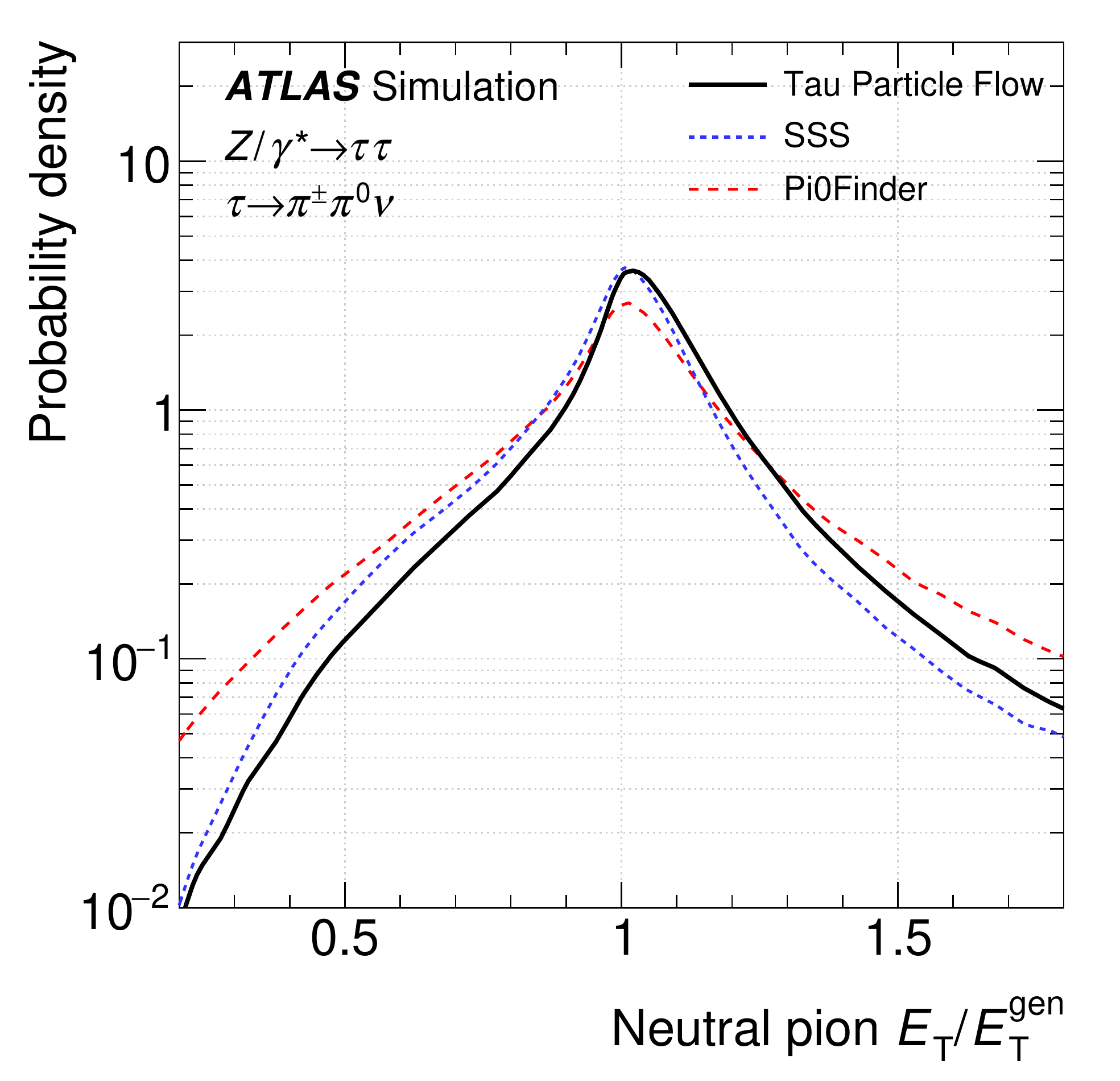}}
\subfigure[\label{subfig:neutralpion_resoeta}]{\includegraphics[width=0.45\columnwidth]{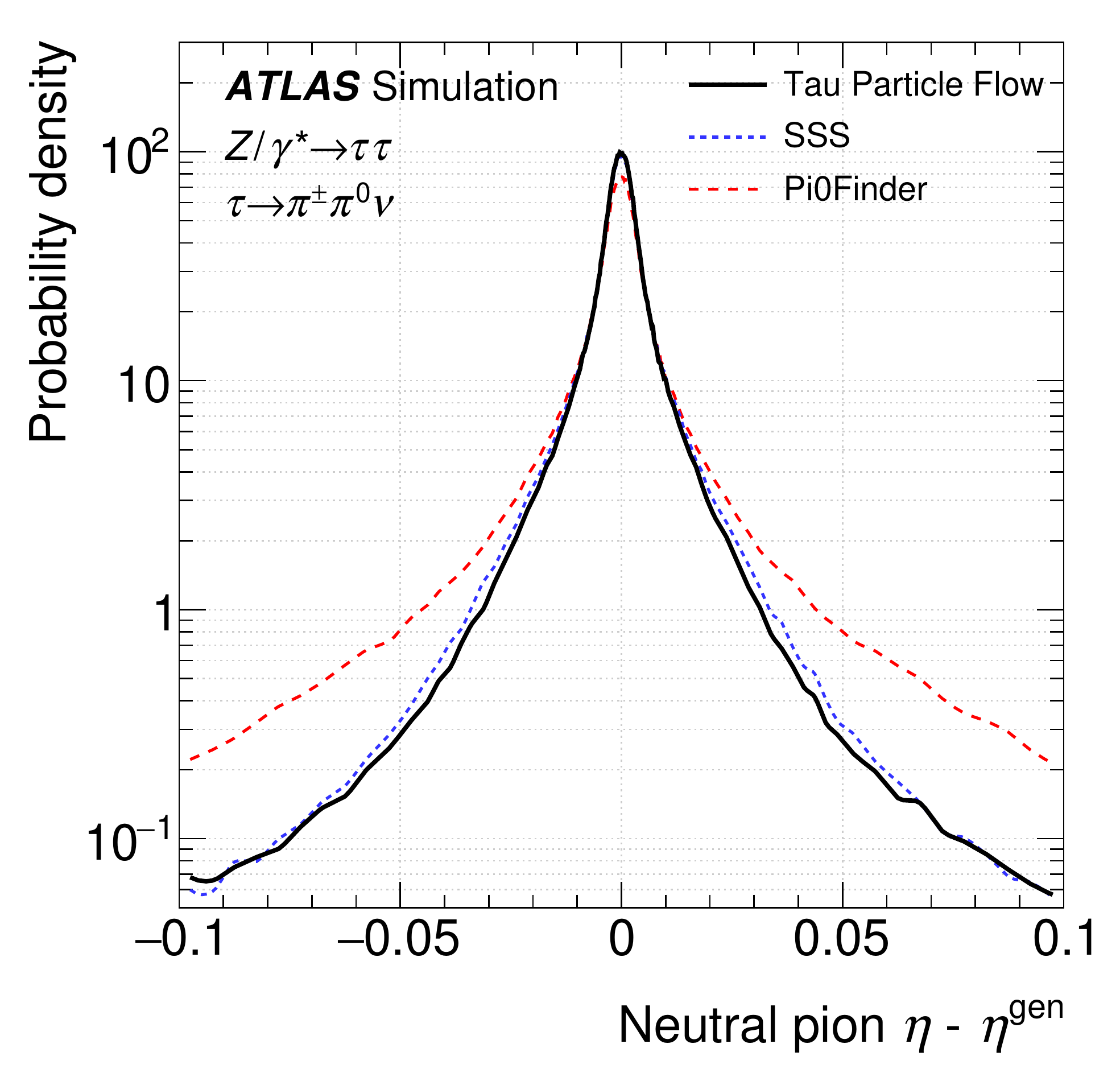}}
\subfigure[\label{subfig:neutralpion_resophi}]{\includegraphics[width=0.45\columnwidth]{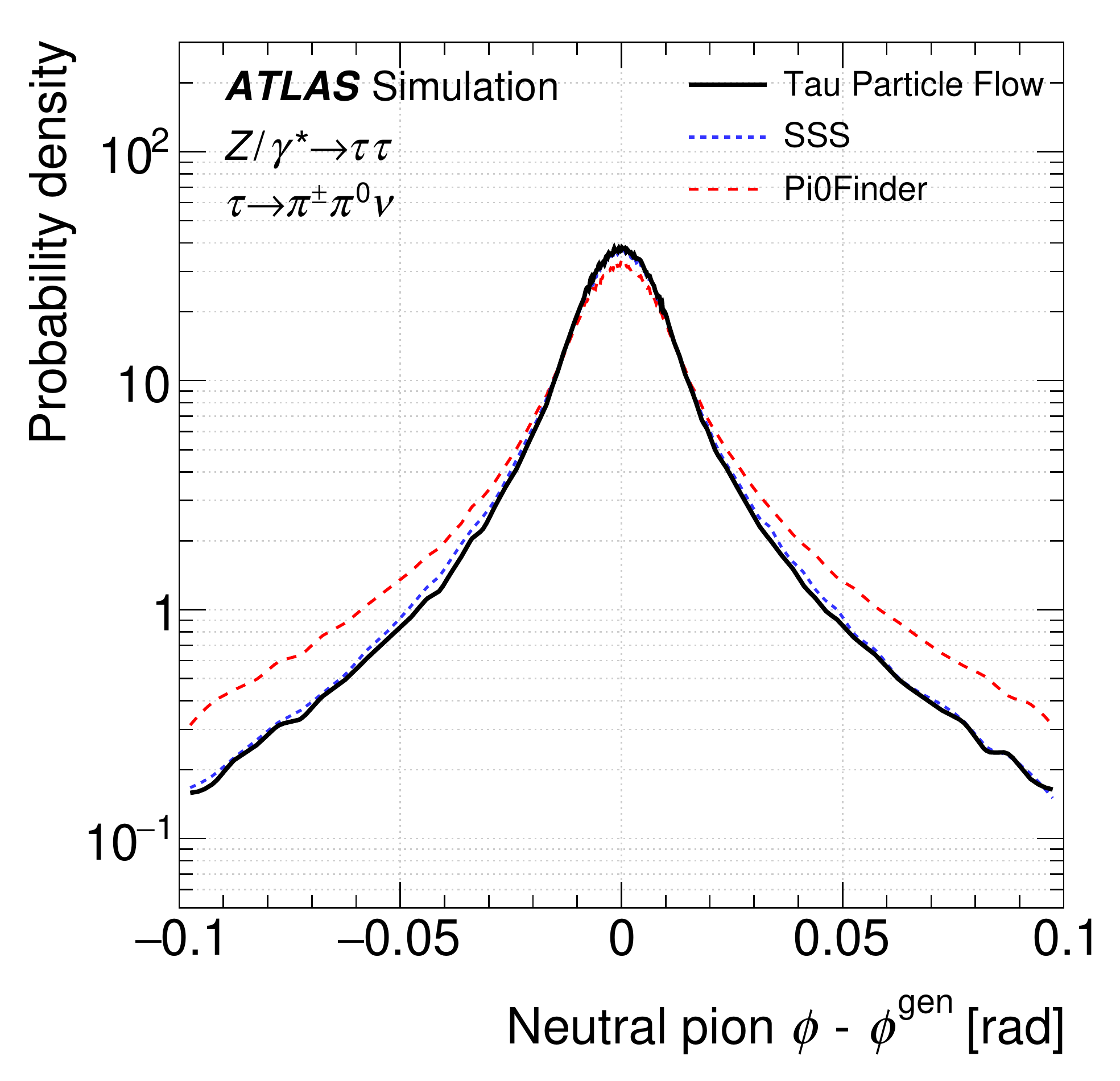}}
\caption{
Distributions of the \pizero residuals in (a) transverse energy \et, (b) pseudorapidity $\eta$ and (c) azimuth $\phi$ in
correctly reconstructed \oneponen decays of tau leptons in simulated 
\Ztautau{} events.
}\label{fig:neutralpion_reso}
\end{center}
\end{figure}

The \pizero \et, $\eta$ and $\phi$ residual distributions for all \pizero
reconstruction algorithms are shown in Figs.~\ref{subfig:neutralpion_resoet},
\ref{subfig:neutralpion_resoeta} and \ref{subfig:neutralpion_resophi},
respectively. The core angular resolutions for each algorithm are quite similar
with $\sim$0.0056 in $\eta$ and $\sim$0.012\,rad in $\phi$.  The \clusterbased
algorithm has the poorest performance, with core resolutions of 0.0086 and
0.016\,rad in $\eta$ and $\phi$, respectively, and significantly larger tails.
The core \et resolutions are almost identical for the \newmethod and
\cellbased, both with 16\%, compared to 23\% for \clusterbased.  The \newmethod
and \cellbased{} both show a shift in the reconstructed \et of a few percent,
due to incomplete subtraction of the \charged remnant. In the calculation of
the \tauhadvis{} four-momentum in the \newmethod
(Section~\ref{sec:fourmomentum}), this bias is corrected for by a
decay-mode-dependent calibration.
Despite the more sophisticated shower subtraction employed in the \cellbased
algorithm, it does not perform significantly better; the improvement in the 
total fraction of correctly classified tau decays
is $\sim$1\%.  This is partly because many of the
\neutrals are sufficiently displaced from \chargeds so that they have little
energy contamination and are unaffected by the subtraction, and partly because
the signature of clusters that contain \pizeros, even in the presence of
overlapping \charged energy, is distinct enough for the BDT to identify. 
 Contributions from \pileup have little effect on the \neutral reconstruction
in \newmethod; on average the \et increases by $\sim$$15\MeV$ and its
resolution degrades fractionally by $\sim$$0.5\%$ per additional reconstructed
vertex. 
\end{sloppypar}

\subsection{Reconstruction of individual photon energy deposits in EM1}
\label{sec:recoshots}

During the \pizero reconstruction, the energy deposits from both photons
typically merge into a single cluster. Furthermore, for \Ztautau events, in
about half of the \onepxn{} decays misclassified as \oneponen{} by the \pizero
reconstruction, at least three of the photons from two \pizeros are grouped into
a single cluster. The fraction increases for higher \tauhadvis \pt due to
the collimation of the tau decay products. The identification of the energy
deposits from individual photons in the finely segmented EM1 layer can be
exploited to improve the \pizero reconstruction, as discussed in the following.

Almost all photons begin to shower by the time they traverse EM1, where they
deposit on average $\sim$30\% of their energy.  In contrast, particles that do
not interact electromagnetically rarely deposit a significant amount of energy
in this layer, making it ideal for the identification of photons. Furthermore,
the cell segmentation in $\eta$ in this layer is finer than the average
photon separation and comparable to the average photon shower width, allowing
individual photons to be distinguished.  

\begin{figure}[htbp]
\begin{center}
\includegraphics[width=0.50\columnwidth]{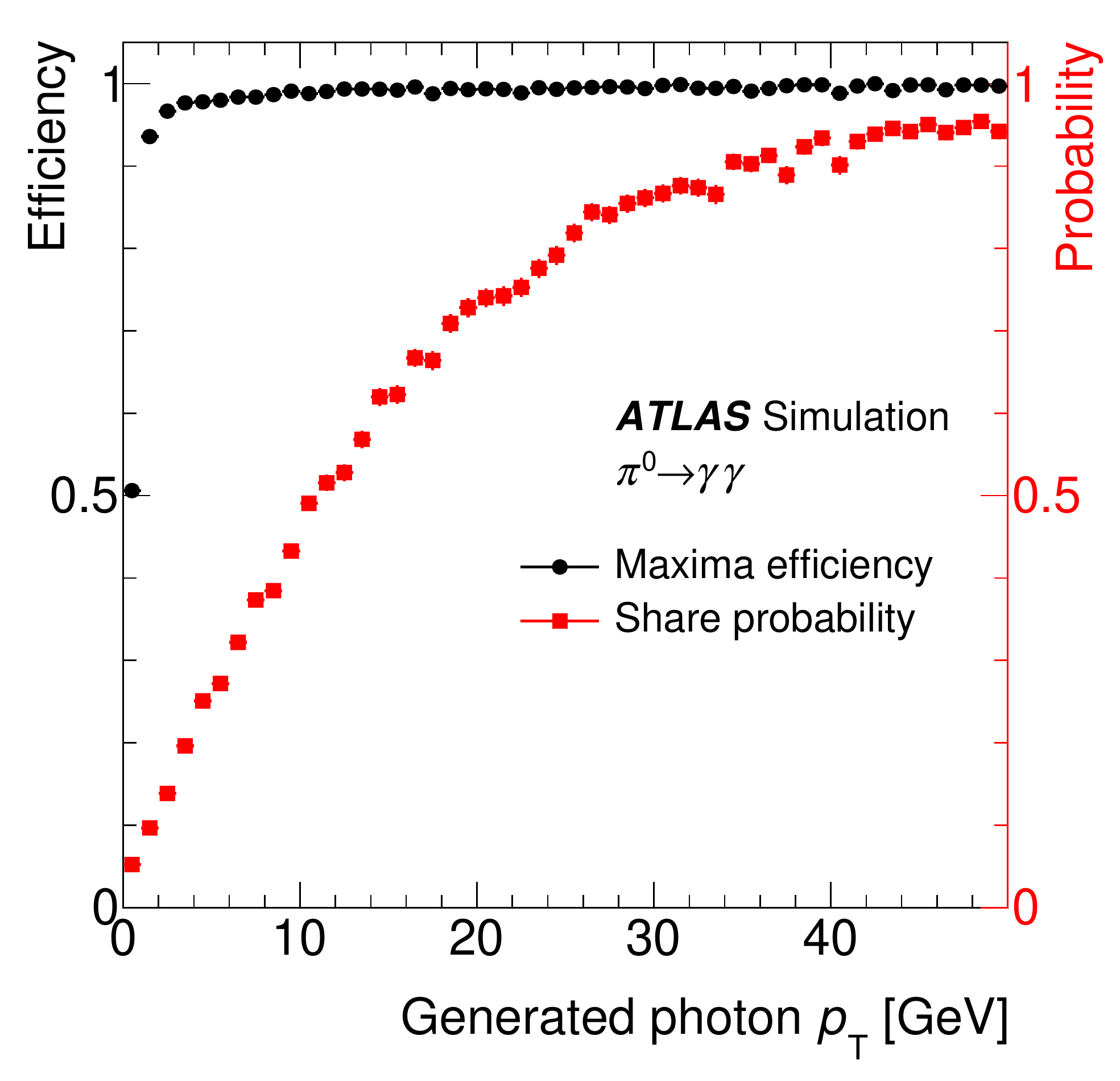}
\caption{
Efficiency for a photon to create a maximum in the first layer 
of the EM calorimeter in simulated $\pi^{0}\to\gamma\gamma$ events
and the corresponding probability to create a maximum that is shared 
with the other photon. The photons are required to not interact 
with the material in the tracking system.
}\label{fig:shoteff}
\end{center}
\end{figure}

The reconstruction of energy deposits in EM1 proceeds as follows.  First, local
energy maxima are searched for within the core region. A local maximum is
defined as a single cell with $\et>100\MeV$ whose nearest neighbours in $\eta$
both have lower \et. Maxima found in adjacent $\phi$ cells are then combined:
their energy is summed and the energy-weighted mean of their $\phi$ positions
is used.  Figure~\ref{fig:shoteff} shows the efficiency for photons to create a
local maximum (maxima efficiency), evaluated in the sample of single \pizeros.
The efficiency decreases rapidly at low photon \pt{} as many of the photons
fall below the $100\MeV$ threshold.  The fraction of misreconstructed maxima
due to noise or fluctuations from the photon shower is very low for maxima with
$\et>500\MeV$, but increases quickly at lower \et.  At high photon \pt{},
corresponding to high \pizero \pt{}, the boost of the \pizero becomes large
enough that the pair of photons almost always creates a single maximum.
Figure~\ref{fig:shoteff} also shows the probability that a maximum is shared
with the other photon in the single \pizero sample (share probability).

The \onepxn{} decay mode classification is improved by counting the number of
maxima associated with \neutrals.  An energy maximum is assigned to a \neutral
if its cell is part of the \neutral cluster and it has an \et of more than
300--$430\MeV$ (depending on the $\eta$ region). The energy threshold is
optimised to maximise the total number of correctly classified tau decays.
Maxima with $\et>10\GeV$ are counted twice, as  they contain the merged energy
deposits of two photons from a \pizero decay with a probability larger than
95\%. Finally, \tauhadvis candidates that were classified as \oneponen{}, but
have a \neutral with at least three associated maxima are reclassified as
\onepxn. The method recovers 16\% of misclassified \onepxn{} decays with a
misclassification of \oneponen{} decays of 2.5\%.

\subsection{Decay mode classification}
\label{sec:recodecaymode}

Determination of the decay mode by counting the number of reconstructed
\chargeds and \neutralIDs alone can be significantly improved
by simultaneously analysing the kinematics of the tau decay products,
the \pizero identification scores and the number of photons from the previous
reconstruction steps. Exploitation of this information is performed via BDTs.

As the most difficult aspect of the classification is to determine the number
of \pizeros, three decay mode tests are defined to distinguish between the
following decay modes: \hpms with zero or one \pizero, \zerovone; \hpms with 
one or more \pizeros, \onevx; and 3\hpms with and without \pizeros, \zerovx.
Which of the three tests to apply to a \tauhadvis candidate is determined as
follows. The \tauhadvis candidates with one or three associated tracks without any
reconstructed \neutrals are always classified as \onepzeron{} or
\threepzeron{}, respectively. The \tauhadvis{} candidates with one associated track
and at least two \neutrals, of which at least one is \neutralID, enter the
\onevx test. The \tauhadvis candidates with one \neutralID that are classified as
\onepxn by counting the photons in this cluster, as described in
Section~\ref{sec:recoshots}, retain their classification and are not considered
in the decay mode tests.  The remaining \tauhadvis candidates with one or three
associated tracks enter the \zerovone or \zerovx tests, respectively.  

A BDT is trained for each decay mode test using \tauhadvis candidates from
simulated \Ztautau events, to separate \tauhadviss of the two generated decay
types the test is designed to distinguish. The \tauhadvis candidates entering each decay mode test are then further
categorised based on the number of \neutralIDs. A threshold is placed on the
output BDT score in each category to determine the decay mode. The thresholds
are optimised to maximise the number of correctly classified \tauhadvis
candidates. The BDT training was not split based on the number of \neutralIDs
due to the limited size of the training sample.
 
The variables used for the decay mode tests are designed to discriminate
against additional misidentified \neutrals, which usually come from imperfect
\hpm subtraction, \pileup or the underlying event. The associated clusters
typically have low energy and a low \pizero identification score. Remnant
clusters from imperfect \charged subtraction are also typically close to the
\charged track and have fewer associated photon energy maxima. If the \neutral
clusters originate from tau decays, their directions and fractional energies
are correlated with each other. Additionally, with increasing number of tau
decay products, the available phase space per decay product becomes smaller.
Each variable used in the BDTs is described briefly in
Table~\ref{tab:recodecaymode:vars}.
Table~\ref{tab:recodecaymode:classification_bdts} summarises the decay mode
tests and indicates which variables are used in each.

\begin{table}
\centering
\begin{tabular}{p{0.80\textwidth}}
\toprule 

{\bf \pizero identification score of the first \neutral, \panpiscoreone}\\
\pizero identification score of the \neutral with the highest \pizero identification score\\[1ex]

{\bf \et fraction of the first \neutral,  \panneutfracone }\\
\et of the \neutral with the highest \pizero identification score, divided by
the \et-sum of all \neutrals and \chargeds\\[1ex]

{\bf Hadron separation,  \panpidr }\\
$\Delta R$ between the \charged and the \neutral with
  the highest \pizero identification score\\[1ex]

{\bf \charged distance,  \panpidist }\\
\et-weighted $\Delta R$ between the \charged and the \tauhadvis axis,
which is calculated by summing the four-vectors of all \chargeds and
\neutrals\\[1ex]

{\bf Number of photons,  \pannumpho }\\
Total number of photons in the \tauhadvis, as reconstructed in
Section~\ref{sec:recoshots}\\[1ex]

{\bf \pizero identification score of second \neutral,  \panpiscoretwo }\\
\pizero identification score of the \neutral with the second-highest
\pizero identification score\\[1ex]

{\bf \neutral \et fraction, \panneutfrac }\\
\et-sum of \neutrals, divided by the \et-sum of \neutrals and \chargeds\\[1ex]

{\bf \neutral mass,  \panneutmass }\\
Invariant mass calculated from the sum of \neutral four-vectors\\[1ex]

{\bf Number of \neutral,  \pannumneut}\\[1ex]

{\bf Standard deviation of the \charged \pt,  \panchargedetvar }\\
Standard deviation, calculated from the \pt values of the \chargeds
for \tauhadvis with three associated tracks\\[1ex]

{\bf \charged mass,  \panchargedmass }\\
Invariant mass calculated from the sum of \charged four-vectors\\[1ex]

\bottomrule

\end{tabular}
\caption{Variables used in the BDTs for the \tauhadvis decay mode
  classification. They are designed to discriminate against additional
  misidentified \neutrals, which usually come from imperfect
  subtraction, \pileup or the underlying event.}
\label{tab:recodecaymode:vars}
\end{table}

\begin{table}[htp] 
\begin{center}

\begin{tabular}{cccc}
\toprule
Decay mode test               & $N(\neutral)$ & $N(\neutralID)$ & Variables \\
\midrule
\multirow{2}{*}{\zerovone} & $\ge1$        & 0               & \multirow{2}{*}{\panpiscoreone, \panneutfracone, \panpidr, \panpidist, \pannumpho} \\
                              & 1             & 1               & \\
\midrule
\multirow{2}{*}{\onevx} & $\ge2$        & 1               & \multirow{2}{*}{\panpiscoretwo, \panneutfrac, \panneutmass, \pannumneut, \pannumpho} \\
                              & $\ge2$        & $\ge2$          & \\
\midrule
\multirow{2}{*}{\zerovx} & $\ge1$        & 0               & \multirow{2}{*}{\panpiscoreone, \panneutfrac, \panchargedetvar, \panchargedmass, \pannumpho} \\
                              & $\ge1$        & $\ge1$          & \\
\bottomrule
\end{tabular}
\caption{
Details regarding the decay mode classification of the \newmethod.
BDTs are trained to distinguish decay modes in three decay mode tests. 
The \tauhadviss entering each test are further categorised based on 
the number of reconstructed, $N(\neutral)$, and identified, $N(\neutralID)$,
neutral pions. The variables used in the BDTs for each test are listed. 
}
\label{tab:recodecaymode:classification_bdts}
\end{center}
\end{table}

\begin{figure}[htbp]
  \begin{center}
    \includegraphics[width=0.50\columnwidth]{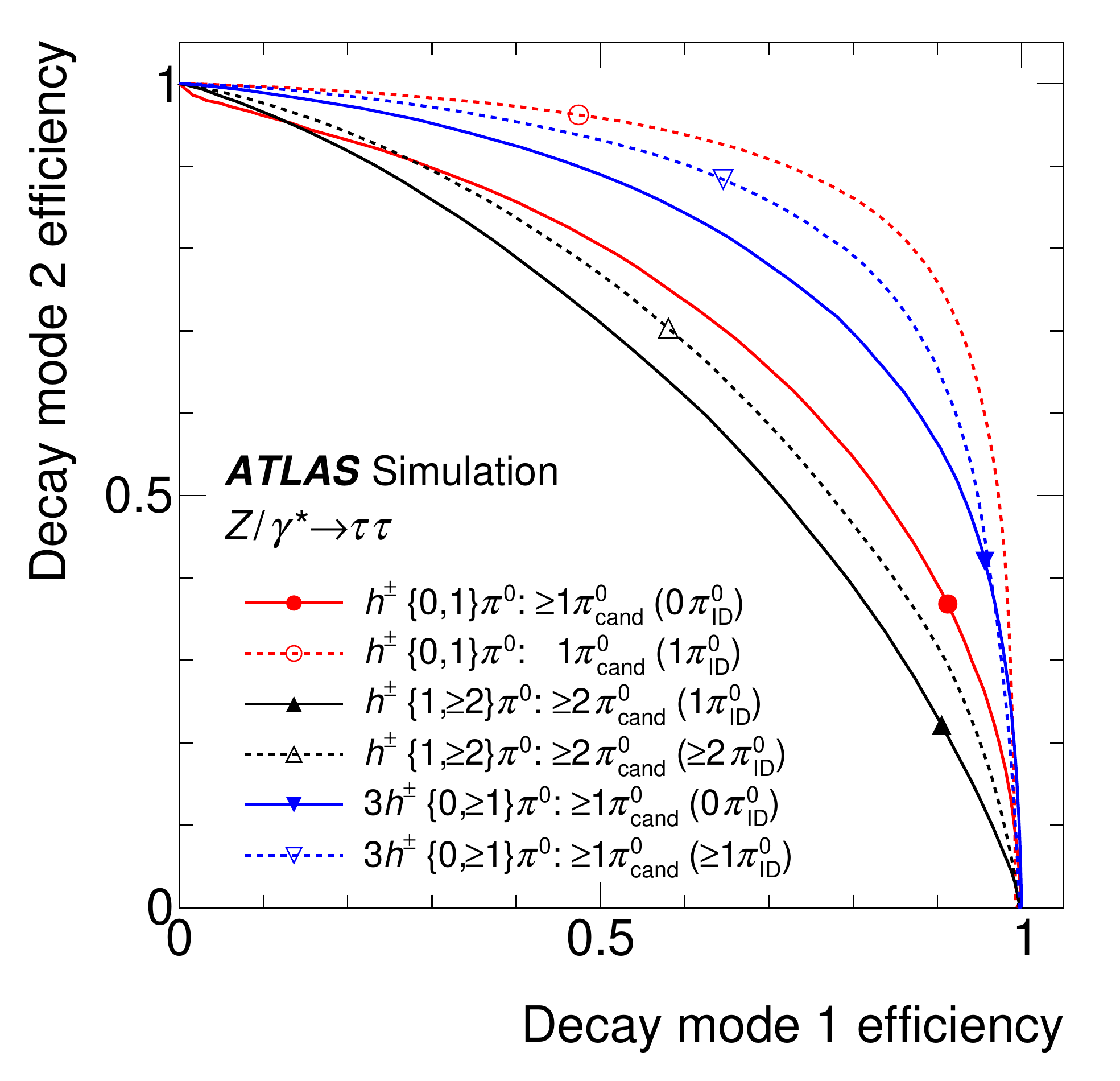}
    \caption{Decay mode classification efficiency for the \zerovone{},
      \onevx{}, and \zerovx{} tests. For each test, ``decay mode 1'' corresponds to the mode 
      with fewer \pizeros.  Working points corresponding to the optimal 
      thresholds on the BDT score for each test are marked. 
      }
  \label{fig:recodecaymode:bdt_distribution}
  \end{center}
\end{figure}

\begin{figure}[htbp]
  \begin{center}
    \includegraphics[width=0.50\columnwidth]{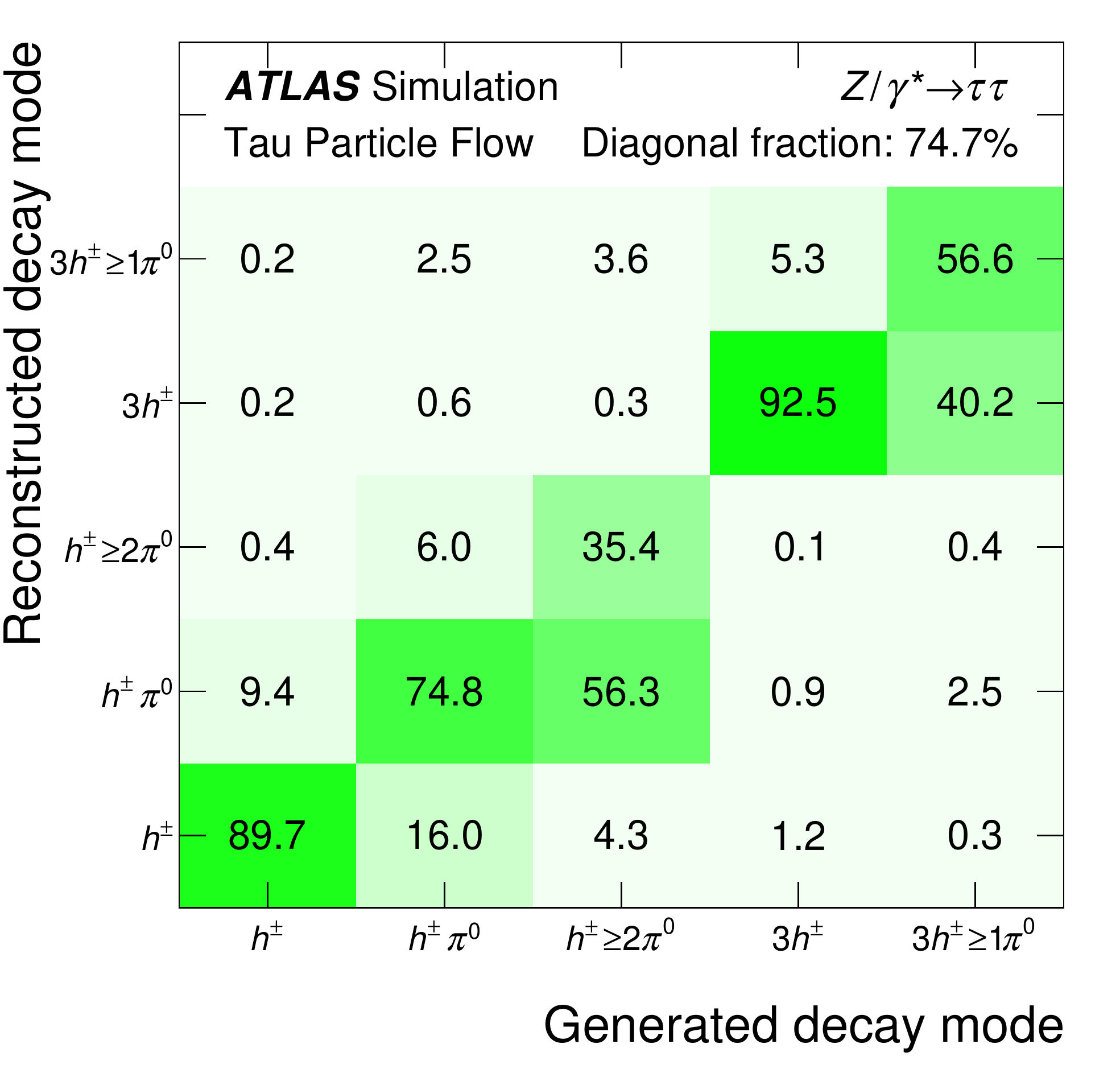}
    \caption{
Decay mode classification efficiency matrix showing the probability for a given
generated mode to be reconstructed as a particular mode by the \newmethod after
final decay mode classification in simulated \Ztautau{} events. Decays
containing neutral kaons are omitted. Only decays from \tauhadviss that are
reconstructed and pass the selection described in
Section~\ref{sec:physicsobjects} are considered.
The statistical uncertainty is negligible. 
    }
  \label{fig:migrationmatrices_eff}
  \end{center}
\end{figure}

\begin{figure}[htbp]
  \begin{center}
    \subfigure[\label{subfig:datamc:ztautau:classification}]{\includegraphics[width=0.49\columnwidth]{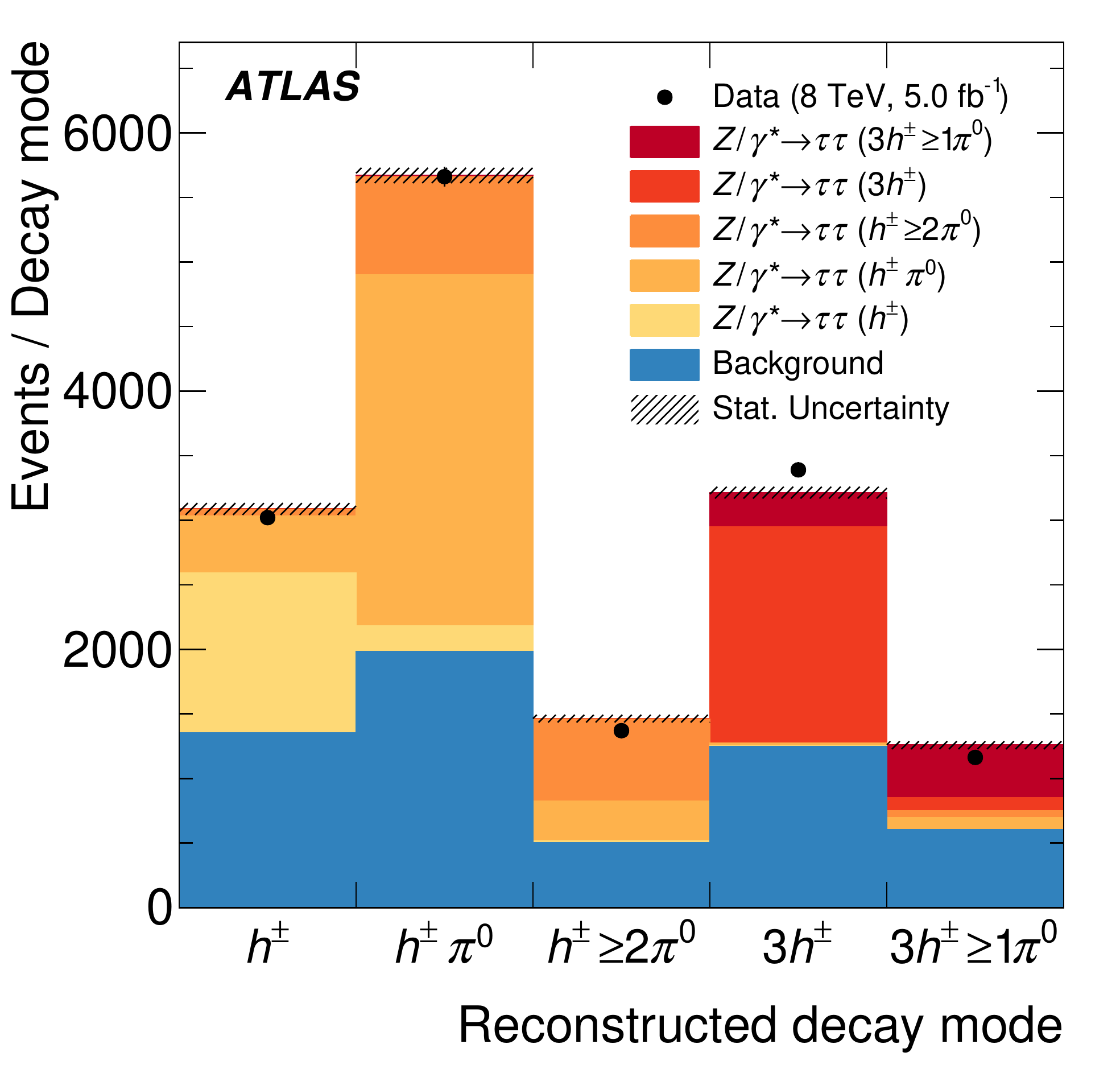}}
    \subfigure[\label{subfig:datamc:zmumu:classification}]{\includegraphics[width=0.49\columnwidth]{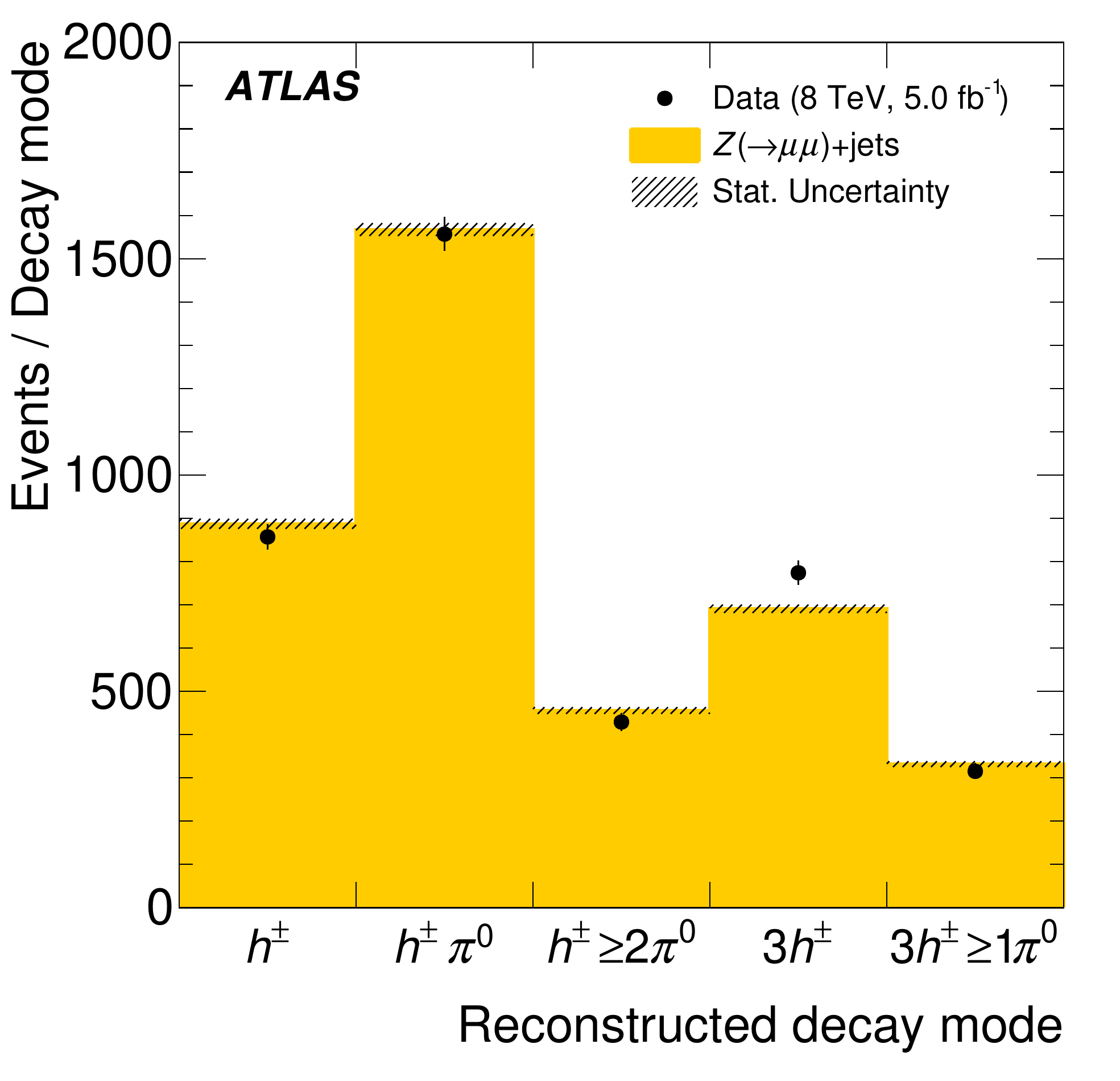}}
    \caption{
      Number of \tauhadvis candidates for each classified decay mode in the (a)
      \Ztautau and the (b) \Zmumujets tag-and-probe analyses.  The simulated \Ztautau
      sample is split into contributions from each generated tau decay mode. The background 
      in the \Ztautau analysis is dominated by multijet and \Wmunujets{} production. 
      The simulated \Zmumujets events are 
      reweighted so that the $Z$ boson \pt distribution 
      and the overall normalisation match that in the data. 
      The hatched band represents the statistical uncertainty on the prediction.
    }
    \label{fig:datamc:ztautau:classification}
  \end{center}
\end{figure}

\begin{sloppypar}
Figure~\ref{fig:recodecaymode:bdt_distribution} shows the discrimination power
of the tests categorised by the number of \neutrals and \neutralIDs. The decay
mode fractions at the input of each test vary strongly, which impacts the
position of the optimal BDT requirements.
The resulting classification matrix is shown in
Fig.~\ref{fig:migrationmatrices_eff}. The total fraction of correctly classified tau decays
is 74.7\%. High efficiencies in the important \onepzeron, \oneponen and
\threepzeron modes are achieved. The {\em decay mode purity} is defined as the
fraction of \tauhadvis candidates of a given reconstructed mode which
originated from a generated \tauhadvis of the same mode, also calculated using
\tauhadviss in simulated \Ztautau events. The purity of the \onepzeron,
\oneponen and \threepzeron decay modes is 70.3\%, 73.5\% and 85.2\%,
respectively. For comparison, in the \baseline reconstruction where \pizero
reconstruction was not available,  the fractions of generated \onepzeron and
\oneponen in \tauhadviss with one reconstructed track are 27.4\% and 52.2\%,
respectively, and the fraction of \threepzeron in \tauhadviss with three
reconstructed tracks is 68.9\%.  Decays containing neutral kaons are omitted
from the table. They are classified as containing \pizeros approximately half
of the time. Contributions from \pileup have little effect on the
classification efficiency, degrading it by $\sim$$0.04\%$ per additional
reconstructed vertex.  The number of \tauhadvis{} candidates for each
classified decay mode is shown in
Fig.~\ref{subfig:datamc:ztautau:classification} for real \tauhadviss from the
\Ztautau tag-and-probe analysis and in
Fig.~\ref{subfig:datamc:zmumu:classification} for jets from the \Zmumujets
tag-and-probe analysis. While systematic uncertainties have not been evaluated,
the figures indicate reasonable modelling of the decay mode classification for
\tauhadviss and jets. In both selections, the \threepzeron efficiency is
slightly underestimated and the \onepxn and \threepxn efficiencies are slightly
overestimated.
\end{sloppypar}

\subsection{Four-momentum reconstruction}
\label{sec:fourmomentum}

\begin{figure}[htbp]
  \begin{center}
      \subfigure[\label{subfig:etares}]{\includegraphics[width=0.49\columnwidth]{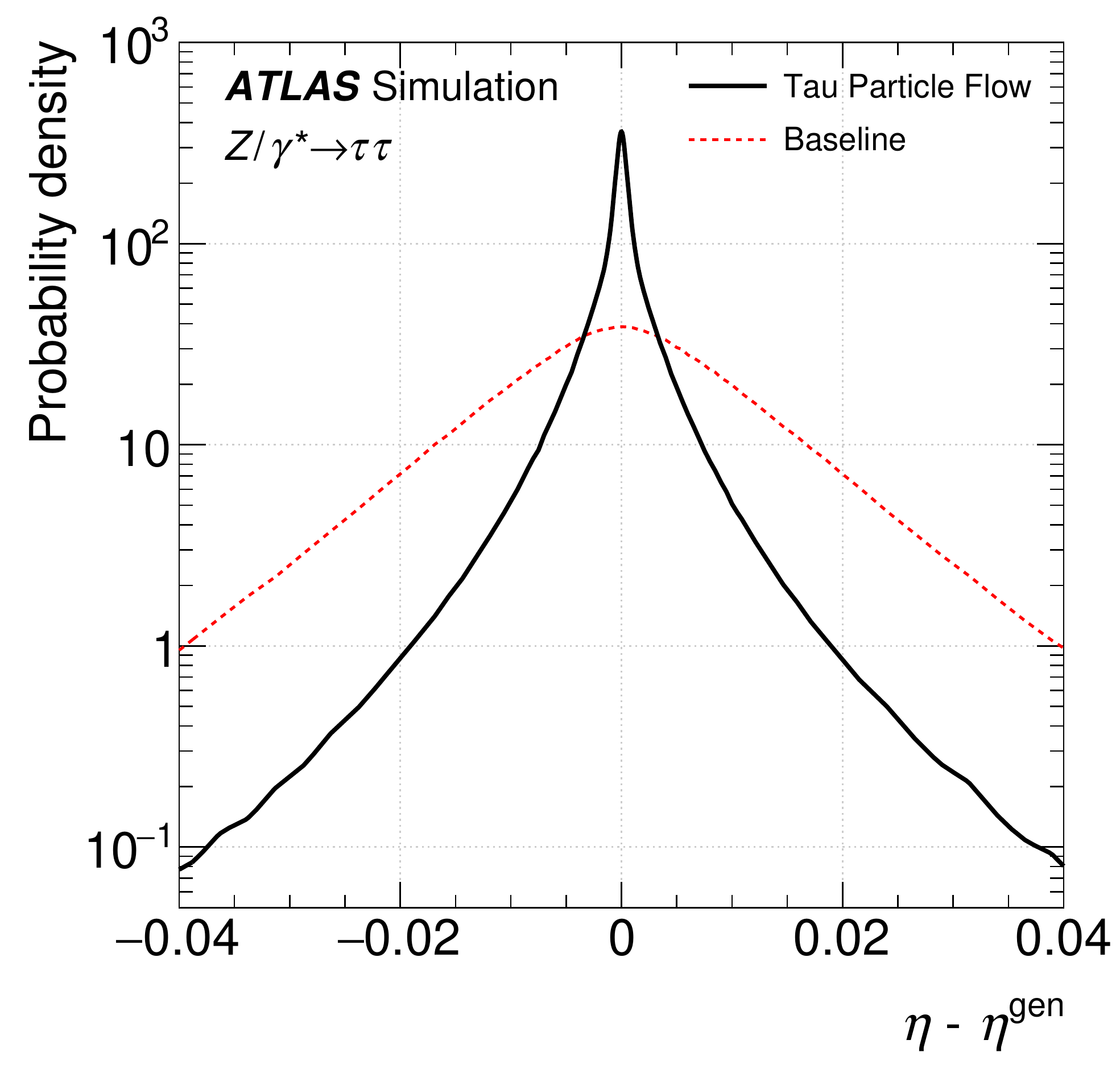}}
      \subfigure[\label{subfig:phires}]{\includegraphics[width=0.49\columnwidth]{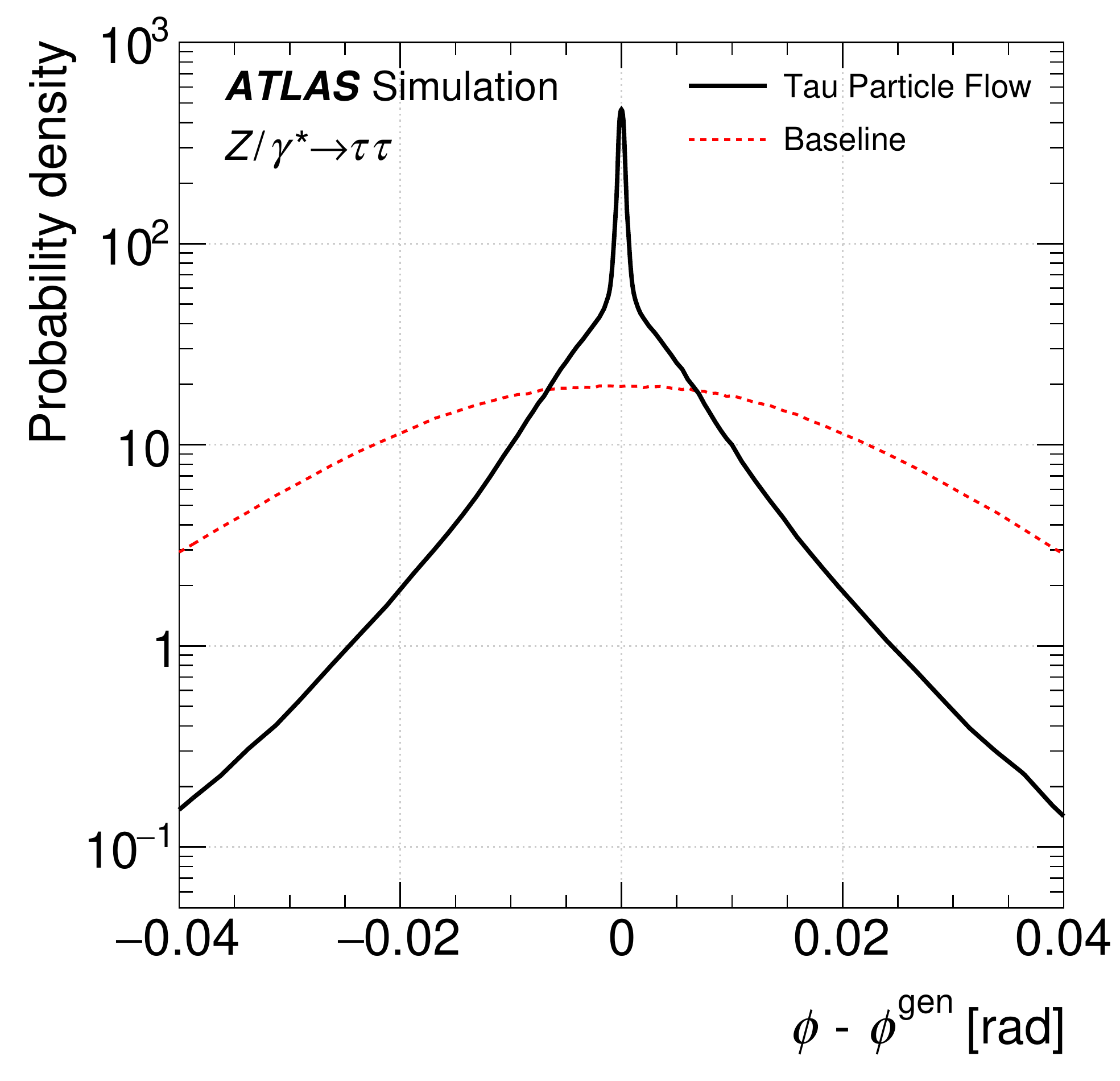}}
    \caption{The \tauhadvis (a) $\eta$ and (b) $\phi$ residual distributions of the \newmethod
      compared to the \baseline 
      reconstruction.}
    \label{fig:angularres}
  \end{center}
\end{figure}

\begin{sloppypar} 
The \tauhadvis four-momentum reconstruction begins with summing the
four-momenta of the \charged and \neutral constituents (\constbased
calculation).  Only the first $n$ \neutrals with the highest \pizero
identification scores are included, where $n$ is determined from the decay mode
classification, and can be at most 2~\neutrals in the \onepxn mode and at most
1~\neutral in the \threepxn mode.  A pion mass hypothesis is used for
\neutrals.  There are two exceptions: if the decay mode is classified as
\oneponen but there are two identified \neutrals, the mass of each is set to
zero and both are added to the \tauhadvis{} four-momentum as they are most
likely photons from a \pizero decay; or if the \tauhadvis{} candidate is
classified as \onepxn because three or more photons are found in a single
\neutral, only this \neutral is added and its mass is set to twice the \pizero
mass.  A calibration is applied to the \constbased \tauhadvis energy in each
decay mode as a function of the \constbased \et{}, to correct for the \neutral
energy bias. The resulting four-momentum is used to set the \tauhadvis
direction in the \newmethod.  Figures~\ref{subfig:etares}
and~\ref{subfig:phires} show distributions of the \tauhadvis{} $\eta$ and
$\phi$ residuals of the \newmethod and the \baseline four-momentum
reconstruction. The core angular resolutions of the \newmethod are 0.002 in
$\eta$ and 0.004\,rad in $\phi$, which are more than five times better than the
\baseline resolutions of 0.012 and 0.02\,rad, respectively.  

\begin{figure}[htbp]
  \begin{center}
      \subfigure[\label{subfig:etres}]{\includegraphics[width=0.45\columnwidth]{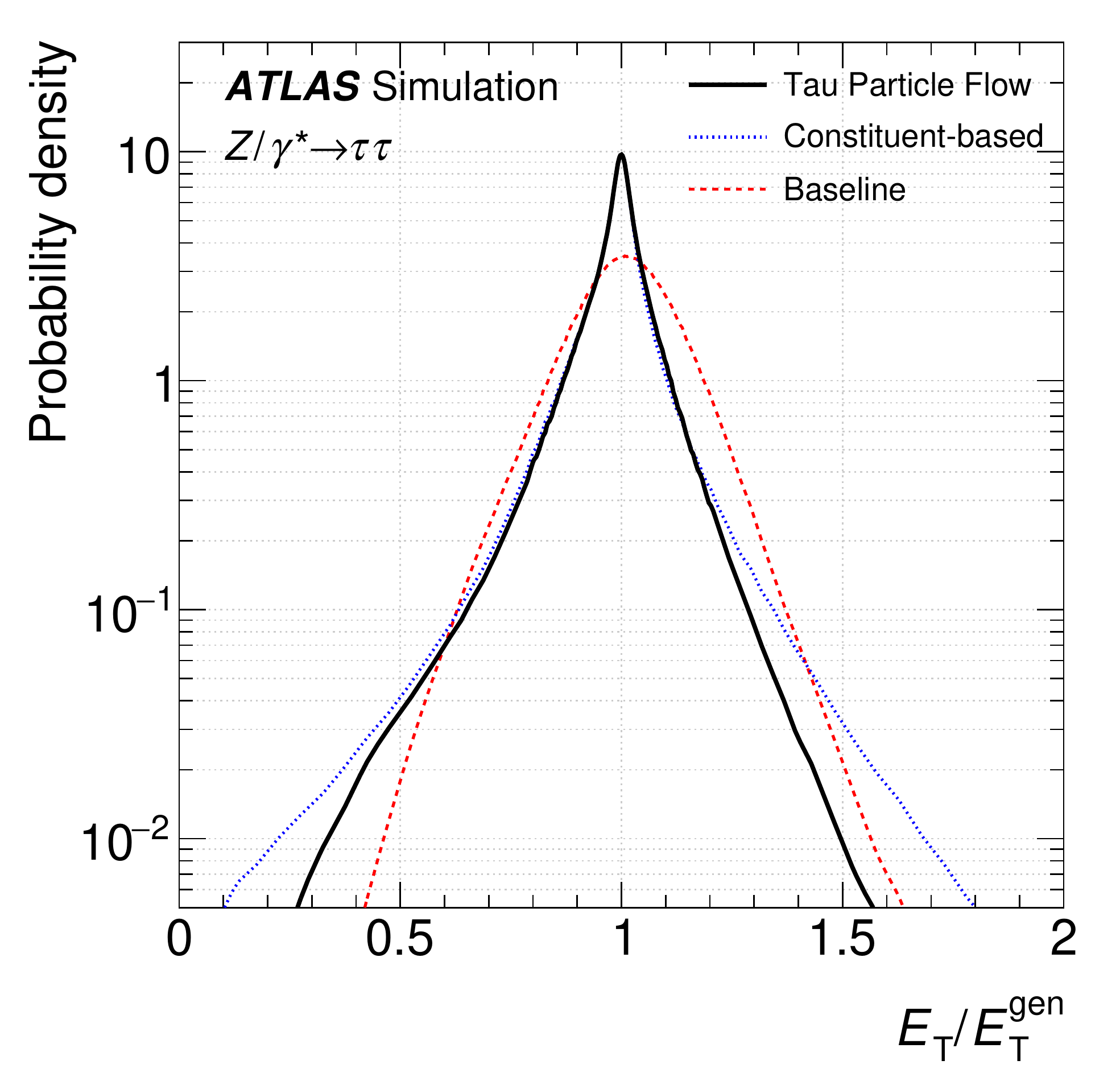}}
      \subfigure[\label{subfig:etresprofile}]{\includegraphics[width=0.45\columnwidth]{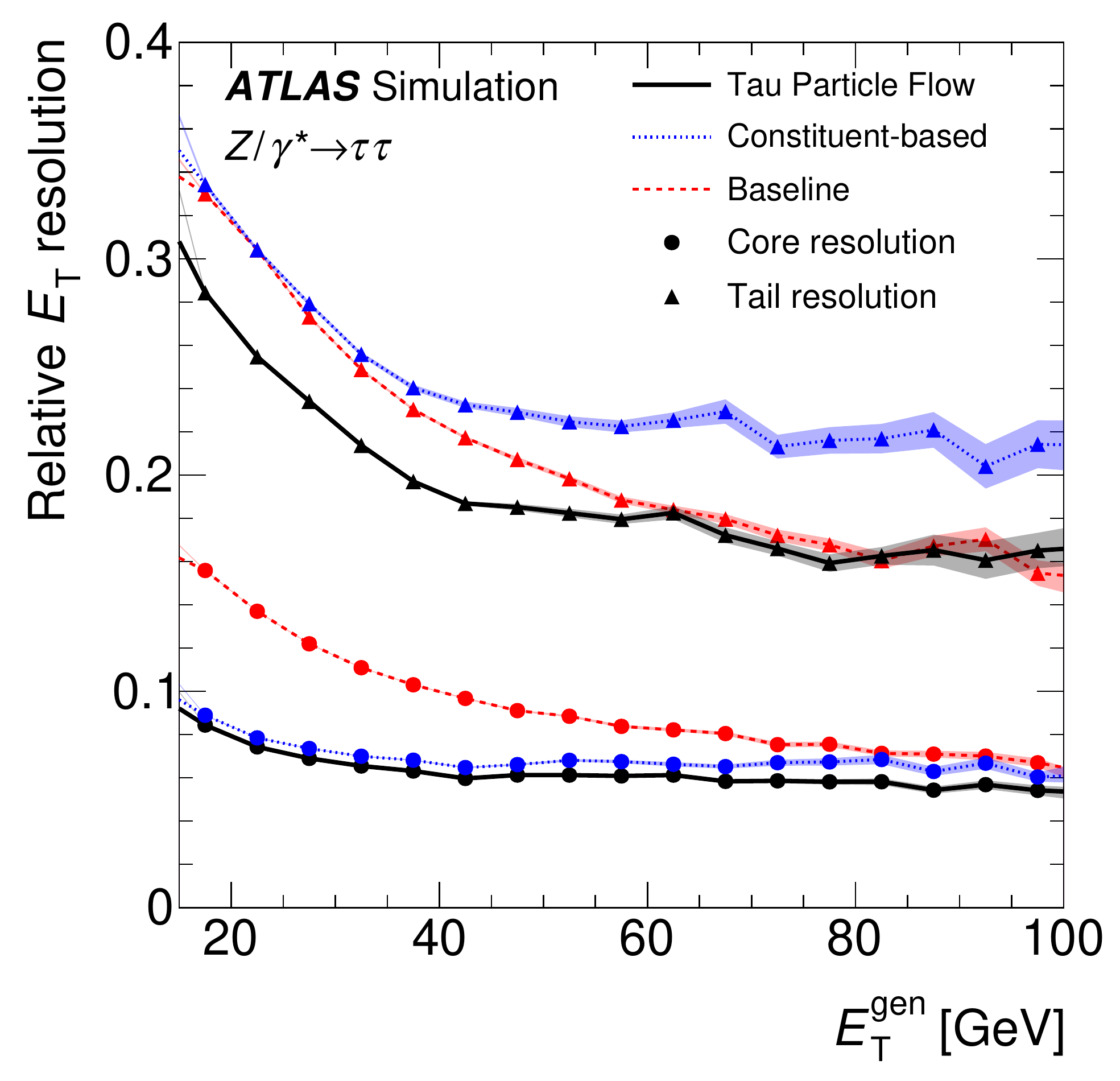}}
      \subfigure[\label{subfig:etresmodes}]{\includegraphics[width=0.45\columnwidth]{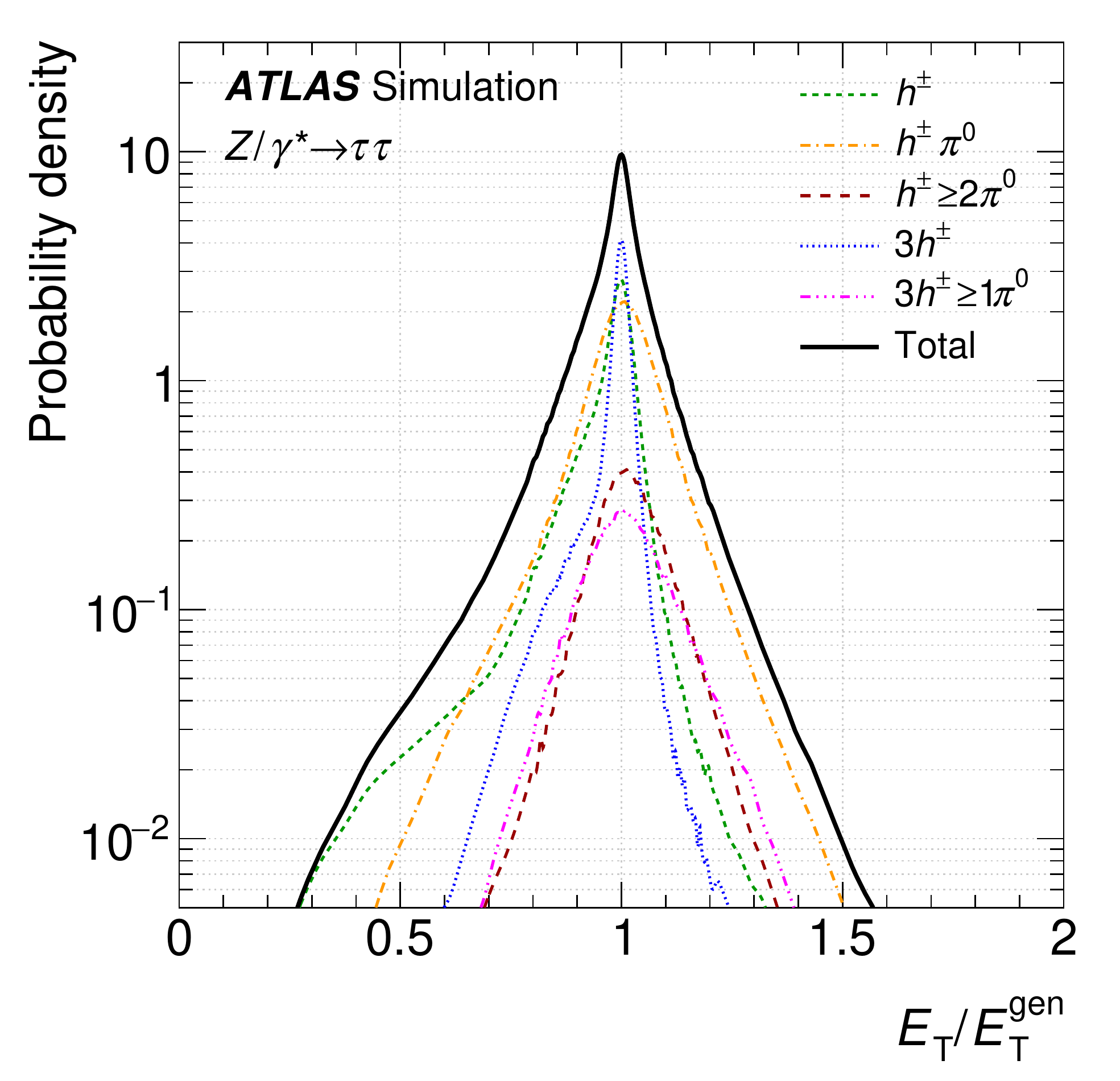}}
    \caption{The (a) \tauhadvis relative \et residual distribution and (b) the
      half-widths spanned by the 68\% and 95\% quantiles, i.e. 
      the core and tail resolutions, of the relative \et
      residual distributions as a function of the generated \tauhadvis \et. The 
      \baseline, 
      \constbased and \newmethod calculations are shown. 
      The relative \et residual distribution of the \newmethod split in the reconstructed 
      decay mode (c) is also shown.
    }
  \label{fig:etres}
  \end{center}
\end{figure}

Figure~\ref{subfig:etres} shows distributions of the \et{} residuals.  The
\constbased calculation is inherently stable against \pileup as both the
decay-mode classification used to select \chargeds and \neutrals, and the
reconstruction of \chargeds and \neutrals themselves, are stable against
\pileup.  The \et increases by $\sim$$6\MeV$ and its resolution degrades
fractionally by $\sim$0.6\% per additional reconstructed vertex.
Figure~\ref{subfig:etresprofile} shows the resolution as a function of the
\et{} of the generated \tauhadvis{}. For the final energy calibration of the
\newmethod, the \constbased \et is combined with the \baseline
\et by weighting each by the inverse-square of their respective
\et-dependent core resolutions, which ensures a smooth transition to high \pt
where the \baseline calibration is superior. The \baseline \et is used if the
two \et values disagree by more than five times their combined core
resolutions, as it has smaller resolution tails.
The resolution of the \newmethod is superior in both the core and tails at low
\et{} with a core resolution of 8\% at an \et of 20~\GeV, compared to 15\% from
the \baseline.  It approaches the \baseline performance at high \et{}.
Contributions from \pileup have little effect on the four-momentum
reconstruction of the \newmethod; the \et increases by $\sim$$4\MeV$ and its
core resolution degrades fractionally by $\sim$0.5\% per additional
reconstructed vertex. 
The \et{} residual distributions of the \newmethod split into the reconstructed
decay modes are shown in Fig.~\ref{subfig:etresmodes}. The total is
non-Gaussian, as it is the sum of contributions with different functional
forms.  Correctly reconstructed decays containing only \chargeds have the best
resolution, followed by correctly reconstructed decays containing \neutrals.
The excellent resolution of these decays leads to a superior overall core
resolution. Misreconstructed decays have the poorest resolution and result in
larger tails. In particular, misestimation of the number of \neutrals leads to
a bias of up to 25\%. Decays containing neutral kaons exhibit a large
low-energy bias because at least some of their energy is typically missed by
the reconstruction.

An alternative method for the \et calibration was also developed, based on
Ref.~\cite{D0Ztautau}. It also uses a combination of calorimetric and
tracking measurements and the \newmethod decay mode classification.  The
\charged \pt is measured using tracks and the \pizero \et is estimated as the
difference between the \et of the seed jet at the EM scale~\cite{LCCalibration}
and the \et from the summed momenta of all \chargeds, scaled by their expected
calorimeter response~\cite{ATL-PHYS-PUB-2014-002}. The method has similar
overall performance to the \newmethod.
\end{sloppypar}

\begin{figure}[htbp]
  \begin{center}
    \subfigure[\label{subfig:datamc:ztautau:vismasssubstr}]{\includegraphics[width=0.49\columnwidth]{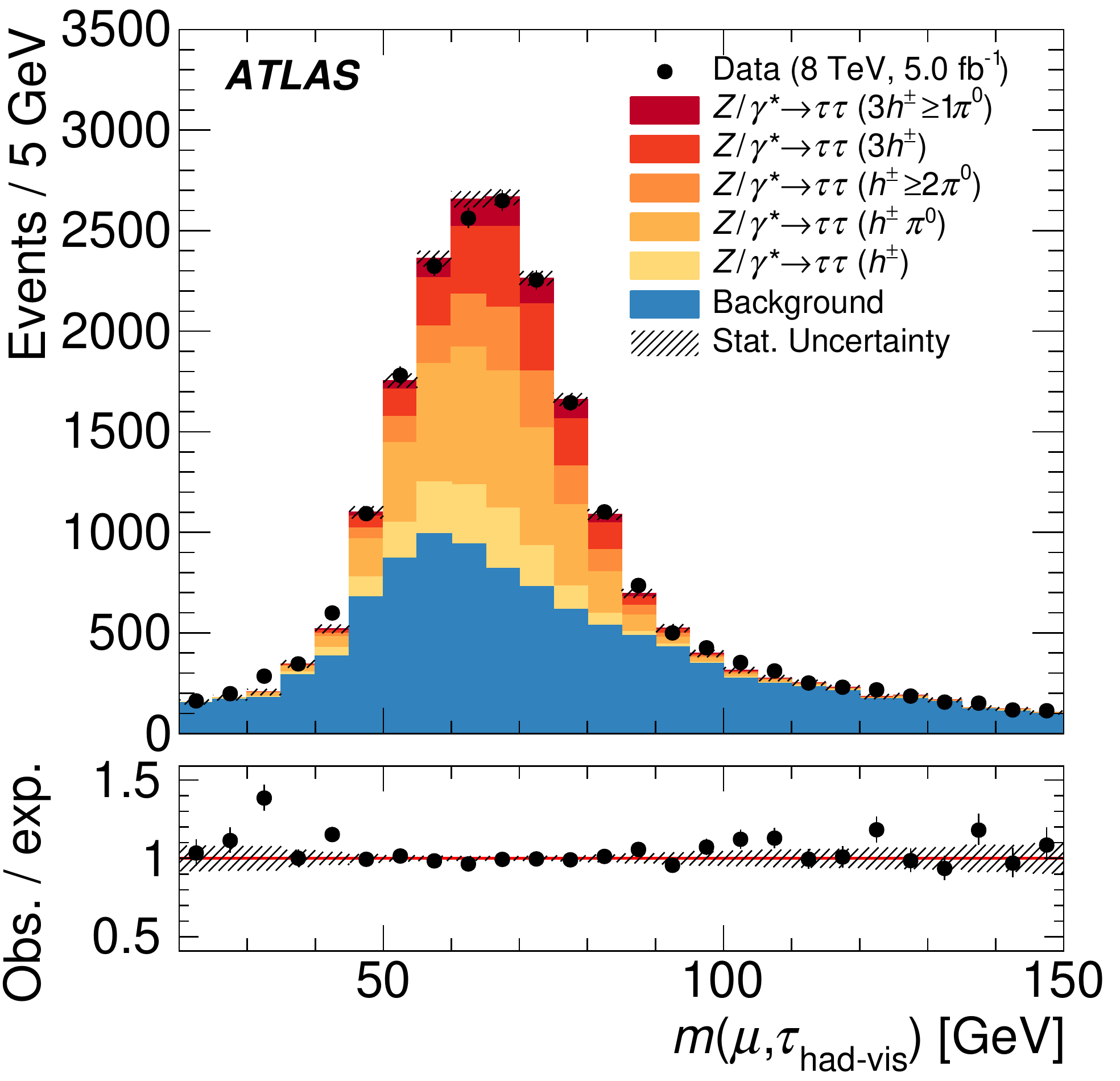}}
    \subfigure[\label{subfig:datamc:ztautau:taumass}]{\includegraphics[width=0.49\columnwidth]{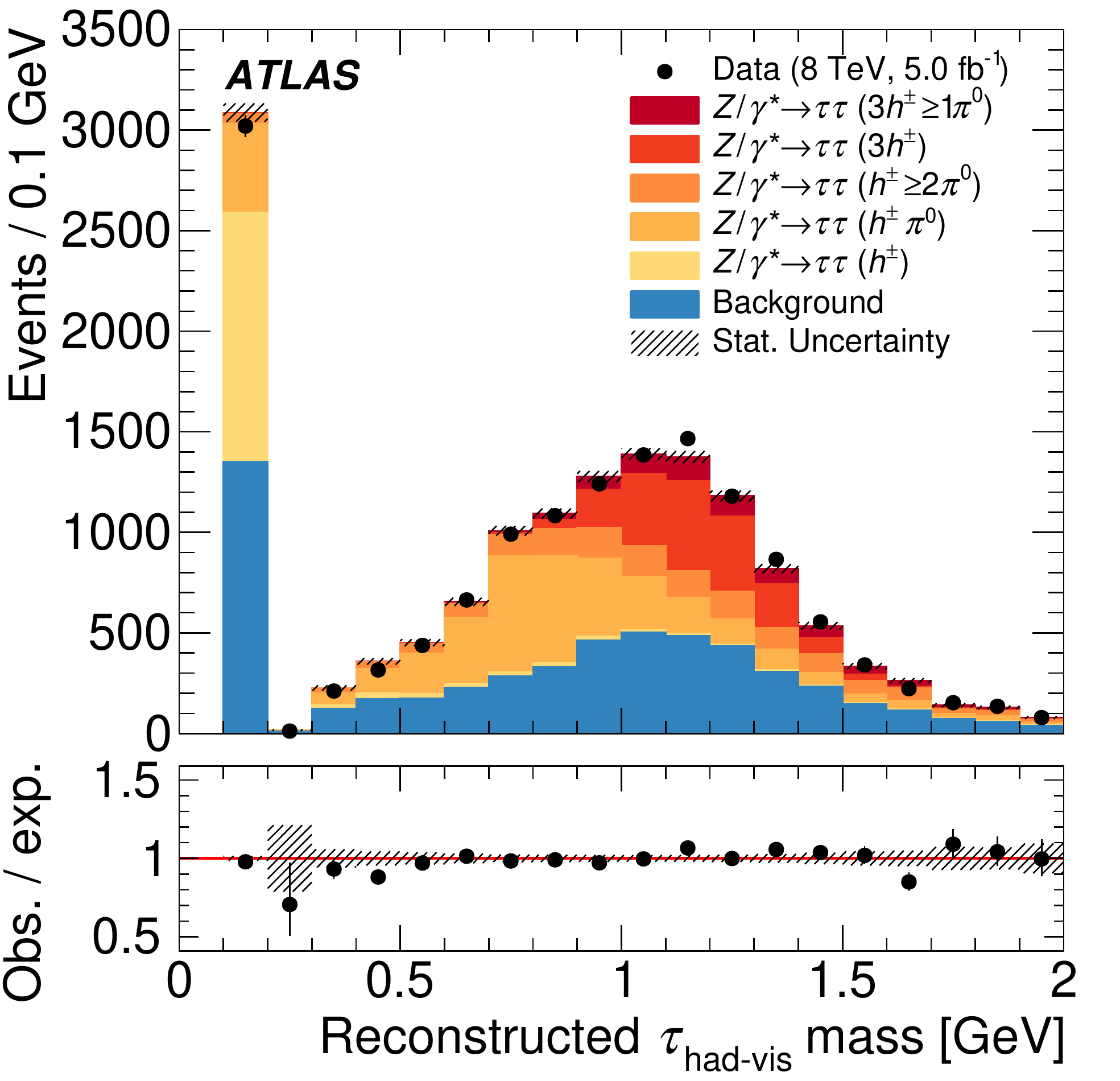}}
      \caption{
        Distribution of (a) the invariant mass of the muon and \tauhadvis, $m(\mu,\tauhadvis)$
        before selection on $m(\mu,\tauhadvis)$ is applied; and (b) the reconstructed mass 
        of the \tauhadvis{},  when using the \newmethod \tauhadvis four-momentum reconstruction 
        in the \Ztautau tag-and-probe analysis.
        The simulated \Ztautau sample is split into contributions from each generated tau decay mode.
        The background is dominated by multijet and \Wmunujets{} production.
        The hatched band represents the statistical uncertainty on the prediction.
      }
    \label{fig:datamc:ztautau:vismasssubstr}
  \end{center}
\end{figure}

Figure~\ref{subfig:datamc:ztautau:vismasssubstr} shows the distribution of the
invariant mass of the muon and \tauhadvis, $m(\mu,\tauhadvis)$, calculated
using the \tauhadvis four-momentum reconstruction from the \newmethod in the
\Ztautau tag-and-probe analysis before selection on $m(\mu,\tauhadvis)$. The
$m(\mu,\tauhadvis)$ has a linear dependence on the \tauhadvis \et and analysis
of the distribution has previously been used to calibrate the \tauhadvis
\et~\cite{Run1TauPaper}.  Data and simulation agree well, indicating that the
\tauhadvis \et is well modelled by the simulation.  
Finally, Fig.~\ref{subfig:datamc:ztautau:taumass} shows the mass spectrum of
the \tauhadvis{} reconstructed with the \newmethod in the \Ztautau
tag-and-probe analysis. The $a_1$ resonance in the \threepzeron mode is
reconstructed with negligible experimental resolution compared to the intrinsic
line shape due to the excellent four-momentum resolution of the inner detector
for \chargeds.  The $\rho$ and $a_1$ resonances in the \oneponen and \onepxn
modes are also visible, but have significant degradation due to the resolution
from the reconstructed \neutral four-momentum. The \tauhadvis mass spectra in
data and simulation agree well, suggesting good modelling of the individual
\charged and \neutral four-momenta. 

\FloatBarrier
\section{Summary and conclusions}
\label{sec:conclusions}

\begin{sloppypar}
This paper presents a new method to reconstruct the individual charged and
neutral hadrons in tau decays with the ATLAS detector at the LHC.  The neutral
pions are reconstructed with a core energy resolution of $\sim$16\%. The
reconstructed hadrons are used to calculate the visible four-momentum of
reconstructed tau candidates and to classify the decay mode, allowing the
decays to be distinguished not only by the number of \chargeds but also by the
number of \pizeros, which is not possible with the existing tau reconstruction.
This improves the purity with which the $\Ptauon\to\piminus\nu$,
$\Ptauon\to\piminus\pizero\nu$ and $\Ptauon\to\piminus\piplus\piminus\nu$
decays can be selected, by factors of 2.6, 1.4 and 1.2, respectively.  The
\tauhadvis core directional resolution is improved by more than a factor of
five and the core energy resolution is improved by up to a factor of two at low
\et{} ($20\GeV$).  The performance was validated using samples of \Ztautau{}
and \Zmumujets{} events selected from $pp$ collision data at $\sqrt{s}=8\TeV$,
corresponding to an integrated luminosity of 5\,\ifb.  The results suggest good
modelling of the \tauhadvis decay mode classification efficiency and
four-momentum reconstruction. 
\end{sloppypar}

\section*{Acknowledgements}
% Acknowledgements for papers with collision data
% Version 7-Feb-2016

% Standard acknowledgements start here
%----------------------------------------------
We thank CERN for the very successful operation of the LHC, as well as the
support staff from our institutions without whom ATLAS could not be
operated efficiently.

We acknowledge the support of ANPCyT, Argentina; YerPhI, Armenia; ARC, Australia; BMWFW and FWF, Austria; ANAS, Azerbaijan; SSTC, Belarus; CNPq and FAPESP, Brazil; NSERC, NRC and CFI, Canada; CERN; CONICYT, Chile; CAS, MOST and NSFC, China; COLCIENCIAS, Colombia; MSMT CR, MPO CR and VSC CR, Czech Republic; DNRF and DNSRC, Denmark; IN2P3-CNRS, CEA-DSM/IRFU, France; GNSF, Georgia; BMBF, HGF, and MPG, Germany; GSRT, Greece; RGC, Hong Kong SAR, China; ISF, I-CORE and Benoziyo Center, Israel; INFN, Italy; MEXT and JSPS, Japan; CNRST, Morocco; FOM and NWO, Netherlands; RCN, Norway; MNiSW and NCN, Poland; FCT, Portugal; MNE/IFA, Romania; MES of Russia and NRC KI, Russian Federation; JINR; MESTD, Serbia; MSSR, Slovakia; ARRS and MIZ\v{S}, Slovenia; DST/NRF, South Africa; MINECO, Spain; SRC and Wallenberg Foundation, Sweden; SERI, SNSF and Cantons of Bern and Geneva, Switzerland; MOST, Taiwan; TAEK, Turkey; STFC, United Kingdom; DOE and NSF, United States of America. In addition, individual groups and members have received support from BCKDF, the Canada Council, CANARIE, CRC, Compute Canada, FQRNT, and the Ontario Innovation Trust, Canada; EPLANET, ERC, FP7, Horizon 2020 and Marie Sk{\l}odowska-Curie Actions, European Union; Investissements d'Avenir Labex and Idex, ANR, R{\'e}gion Auvergne and Fondation Partager le Savoir, France; DFG and AvH Foundation, Germany; Herakleitos, Thales and Aristeia programmes co-financed by EU-ESF and the Greek NSRF; BSF, GIF and Minerva, Israel; BRF, Norway; the Royal Society and Leverhulme Trust, United Kingdom.

The crucial computing support from all WLCG partners is acknowledged
gratefully, in particular from CERN and the ATLAS Tier-1 facilities at
TRIUMF (Canada), NDGF (Denmark, Norway, Sweden), CC-IN2P3 (France),
KIT/GridKA (Germany), INFN-CNAF (Italy), NL-T1 (Netherlands), PIC (Spain),
ASGC (Taiwan), RAL (UK) and BNL (USA) and in the Tier-2 facilities
worldwide.
%----------------------------------------------

\printbibliography

\newpage % ATLAS Collaboration author list
% Data extracted on 29-Oct-2015 for paper reference PERF-2014-06
%\documentclass[11pt]{article}
%\usepackage{a4wide}\begin{document}
\begin{flushleft}
{\Large The ATLAS Collaboration}

\bigskip

G.~Aad$^\textrm{\scriptsize 85}$,
B.~Abbott$^\textrm{\scriptsize 113}$,
J.~Abdallah$^\textrm{\scriptsize 151}$,
O.~Abdinov$^\textrm{\scriptsize 11}$,
R.~Aben$^\textrm{\scriptsize 107}$,
M.~Abolins$^\textrm{\scriptsize 90}$,
O.S.~AbouZeid$^\textrm{\scriptsize 158}$,
H.~Abramowicz$^\textrm{\scriptsize 153}$,
H.~Abreu$^\textrm{\scriptsize 152}$,
R.~Abreu$^\textrm{\scriptsize 116}$
% ...
\bigskip
\\
% ...
$^{11}$ Institute of Physics, Azerbaijan Academy of Sciences, Baku, Azerbaijan\\
% ...
$^{85}$ CPPM, Aix-Marseille Universit{\'e} and CNRS/IN2P3, Marseille, France\\
% ...
$^{90}$ Department of Physics and Astronomy, Michigan State University, East Lansing MI, United States of America\\
% ...
$^{107}$ Nikhef National Institute for Subatomic Physics and University of Amsterdam, Amsterdam, Netherlands\\
% ...
$^{113}$ Homer L. Dodge Department of Physics and Astronomy, University of Oklahoma, Norman OK, United States of America\\
% ...
$^{116}$ Center for High Energy Physics, University of Oregon, Eugene OR, United States of America\\
% ...
$^{151}$ Institute of Physics, Academia Sinica, Taipei, Taiwan\\
$^{152}$ Department of Physics, Technion: Israel Institute of Technology, Haifa, Israel\\
$^{153}$ Raymond and Beverly Sackler School of Physics and Astronomy, Tel Aviv University, Tel Aviv, Israel\\
% ...
$^{158}$ Department of Physics, University of Toronto, Toronto ON, Canada\\
% ...
\end{flushleft}

%\end{document}
% Created with xml2latex.py

\end{document}